\documentclass[structabstract]{aa}

\usepackage{rotating}
\usepackage{color}
\usepackage{graphicx}
\usepackage{aalongtable}
\usepackage{lscape}
\usepackage{txfonts}
 \usepackage{subfig}
 \usepackage{natbib}
\bibpunct{(}{)}{;}{a}{}{,}
\begin{document}

\title{Proper motion survey and kinematic analysis of the $\rho$~Ophiuchi embedded cluster
\thanks{Based on observations collected at the European Southern
Observatory, Chile (64.I-0197, 67.C-0349, 69.C-0230, 71.C-0028, 73.C-0022, 83.D-0635).}
\thanks{Tables 6, 7, and 8 are only available in electronic form at the CDS via anonymous ftp to cdsarc.u-strasbg.fr (130.79.128.5) or via http://cdsweb.u-strasbg.fr/cgi-bin/qcat?/A+A/}}

\author{C. Ducourant \inst{1,2}
  \and R. Teixeira \inst{2,1}
  \and A. Krone-Martins \inst{3,1}
  \and S. Bontemps\inst{1}
  \and D. Despois \inst{1}
  \and P. A. B. Galli \inst{2}
  \and H. Bouy \inst{4}
  \and J.F. Le Campion \inst{1}
  \and M. Rapaport \inst{1}
  \and J.C. Cuillandre \inst{5}
 }

\offprints{ducourant@obs.u-bordeaux1.fr}
\institute{
Laboratoire d'astrophysique de Bordeaux, Univ. Bordeaux, CNRS, B18N, all\'ee Geoffroy Saint-Hilaire, 33615 Pessac, France.
\and
Instituto de Astronomia, Geof\'isica e Ci\^encias Atmosf\'ericas, Universidade de S\~ao Paulo, Rua do Mat\~ao, 1226 - Cidade Universit\'aria, 05508-900 S\~ao Paulo - SP, Brazil.
\and
CENTRA/SIM, Faculdade de Ci\^encias, Universidade de Lisboa, Ed. C8, Lab. 8.5.02, Campo Grande, 1749-016, Lisboa, Portugal.
\and
Centro de Astrobiolog\'\i a, Dpto de Astrof\'\i sica, INTA-CSIC, PO BOX 78, E-28691, ESAC Campus, Villanueva de la Ca\~nada, Madrid, Spain.
\and
CEA/IRFU/SAp, Laboratoire AIM Paris-Saclay, CNRS/INSU, Universit\'e Paris
Diderot, Observatoire de Paris, PSL Research University, F-91191
Gif-sur-Yvette Cedex, France.}

\date{Received 16 October 2015 / Accepted 5 September 2016}

\abstract 
{The $\rho$~Ophiuchi molecular complex and in particular the Lynds L1688 dark cloud  is unique in its proximity ($\sim$130~pc), 
in its richness in young stars and protostars, and in its
youth (0.5~Myr). It is certainly one of the best targets currently accessible
from the ground to study the early phases of star-formation. Proper motion analysis is a very efficient tool for separating members of clusters from field stars, but very few proper motions are available in the $\rho$ Ophiuchi region since most of the young sources are deeply embedded in dust and gas.} 
 {We aim at performing a kinematic census of young stellar objects (YSOs) in the  $\rho$ Ophiuchi F core  and partially in the E core of the L1688 dark cloud.}
 {We run a proper motion program at the ESO New Technology Telescope (NTT) with the Son of ISAAC (SOFI) instrument  over nine years in the near-infrared. We complemented these observations with various public image databases to enlarge the time base of observations and the field of investigation to 0.5$\degr \times 0.5\degr$. We derived positions and proper motions for  2213 objects. From these,  607 proper motions were derived from SOFI observations with a $\sim$1.8 mas/yr accuracy while the remaining objects were measured only from auxiliary data with a mean precision of about $\sim$3 mas/yr.}
 {We performed a kinematic analysis of the most accurate proper motions derived in this work, which allowed us to separate cluster members from field stars and to derive the mean properties of the cluster. From the kinematic analysis we derived a list of 68 members and 14 candidate members, comprising 26 new objects with a high membership probability. These new members are generally fainter than the known ones. We measured a mean proper motion of ($\mu_{\alpha}\cos\delta$, $\mu_{\delta}$)=(-8.2, -24.3)$\pm$0.8 mas/yr for the L1688 dark cloud. A supervised classification was applied to photometric data of members to allocate a spectral energy distribution (SED) classification to the unclassified members.}
 {We kinematically confirmed that the 56 members that were known from previous studies of the $\rho$~Ophiuchi F cluster and that were also part of our survey are members of the cluster, and we added 26 new members. We defined the evolutionary status of the unclassified members of the cluster. We showed that a large part (23) of these new members are probably brown dwarfs, which multiplies the number of known substellar objects in the cluster by a  factor of 3.3.}
\titlerunning{Kinematics in $\rho$ Ophiuchi}
\authorrunning{Ducourant C. et al.}

\maketitle

\section{Introduction}
Stars are believed to form predominantly in groups that gradually lose their content in gas and disperse.  Clusters are  groups that remain stable against tidal disruption by the Galaxy or by passing interstellar clouds and that do not loose their members rapidly (evaporation time $> $100 Myr), whereas associations are looser, less stable groups \citep{Lada(2003)}.
The youngest clusters and associations keep a fresh record of the physical and kinematic conditions involved in their formation. They are the precursors of the visible open clusters and of the dispersed galactic streams of stars.

The youngest of the very young ($< 5$ Myr) clusters are still embedded in their parent cloud. Their properties  and scientific interest in them have been reviewed by \cite{Lada(2003)} who listed 76 clusters. Since these clusters are extremely young, they should not have lost any stars yet because of the gravitational well of the embedding interstellar cloud. As in addition substellar objects have their peak luminosity in their first Myr, embedded clusters constitute excellent benchmarks for studies of the initial mass function, especially at the low-mass end.
Strong absorption ($A_V \sim$ 10-50) by the remaining interstellar matter prevents studying embedded clusters at visible wavelengths, however. Near-infrared (NIR) observations are required and the NIR range is also best suited to studying the spectral energy distribution (SED) of the low-mass objects.

Separating true brown dwarfs from field objects is not simple, however. The confirmed membership of objects to a forming cluster makes their identification as  brown dwarfs much more reliable when it is joined to colour-magnitude and colour-colour diagrams because it fixes their distance and age within a narrow range. This method also produces the least biased samples in this mass range \citep{Luhman(2012b)}. Another debated question is the age determination of the youngest clusters and the possible age spread within a given cluster. Here secure membership is also crucial, as interloping  field stars, especially at the very faint end of the luminosity distribution, can lead to misinterpretations (e.g. \cite{Soderblom(2014)}.

The only membership analysis in clusters that does not depend on a hypothesis is the kinematic analysis which uses the proper motions of objects to separate cluster members from field stars. The main limitations of this method are the accuracy of the proper motions, the limiting magnitude of the observations, and obviously the distance of the cluster. 
Astrometric surveys of embedded clusters are still rare, however mainly because traditionally astrometry is performed at optical wavelengths and no good astrometric reference catalogue exists in the near IR, especially in K band. 
Even the astrometric space mission Gaia \citep{Perryman(2001),Mignard(2008),Lindegren(2010)} will not be able to provide proper motions in such obscure regions, and the ground-based astrometric works in the NIR therefore remain very unique.

A variety of proper motion studies, including surveys of large regions have been conducted in the NIR  (e.g. recently \cite{Dawson(2014)} with  UKIDSS, \cite{Vrba(2004), Vrba(2012)} with USNO, \cite{Pena(2015), Pena(2016)},  \cite{Bouy(2013), Bouy(2015)}). The Upper Sco region is part of the Sco OB2 association, is probably physically related to the $\rho$ Ophiuchi region but somewhat older ($\sim 10$ Myr), is no longer embedded in its parent cloud and has low absorption ($A_V \sim$ 1--2). It has recently been the target of proper motion studies in the NIR (\cite{Luhman(2012a)}, \cite{Lodieu(2013)}). Very few studies of high-absorption regions ($A_V\sim$10-50) have been conducted because they require measuring positions in the less absorbed K band (2.2 $\mu$m; $A_K/A_V\sim$0.1). A very special case is the Galactic centre region ($A_V\sim$27; e.g. \cite{Eckart(2013)}, \cite{Fritz(2010),Fritz(2016)}; \cite{Do(2013),Boehle(2016)}), where very high precision astrometry in K band has been conducted, in particular to study stellar orbits around the central black hole. A wide-field study of the Carina region has also been conducted in K band by \cite{Preibisch(2014)} with proper motion accuracy $\sigma_{\mu} \sim$ 9--10 mas/yr. 

To improve our knowledge of the cluster membership,
we decided to run a NIR proper motion program in the $\rho$ Ophiuchi complex. The $\rho$~Ophiuchi IR cluster is unique in its proximity (120-140~pc) \citep{Wilking(2008)}, in its richness in young stars and protostars, and in its youth because it is one of the youngest known clusters at $\sim$0.5~Myr \citep{Bontemps(2001), Andre(2007)}. It is certainly one of the best targets currently accessible for star-formation studies.

Many works in the $\rho$~Ophiuchi  region tried to assess the membership to the $\rho$~Ophiuchi  complex through spectroscopy and photometry. In a pioneer work, \cite{Bontemps(2001)} performed a census of the population of young stars with IR excesses in this region using Infrared Space Observatory (ISO) \citep{Cesarsky(1996)} observations. \cite{Evans(2003)} delivered the c2d Spitzer \citep{Gallagher(2003)} final data release (DR4) providing observations and characterisation of sources from molecular cores to planetary disks in the mid- to far-infrared wavelengths. In a recent series of paper, \cite{Alves(2010), Alves(2012), Alves(2013)} established which objects belong to the low-mass population in the $\rho$ Ophiuchi molecular cloud through photometric observations in the NIR regime. With a spectroscopic follow-up, the authors characterised the brown dwarf population of an exhaustive list of candidates and detected disks around several brown dwarfs. All these works performed a census of the $\rho$ Ophiuchi complex, that relied on photometric data and extinction models.

We present here
a study of the proper motions in the direction of the sub-cluster $\rho$~Ophiuchi~F of the Lynds L1688 dark cloud that exhibits a clear association of young stars with the Ophiuchi~F molecular dense core. A study recently published by \citep{Wilking(2015)} also addressed proper motions in the $\rho$ Oph cluster (cores B-2,F,C-S,E and A-3, which partly  overlaps the region covered here). This study used the same telescope and ASTROCAM as the USNO IR astrometry program (e.g. \cite{Vrba(2004), Vrba(2012)}) and is complementary to our work. By choosing to derive relative proper motions of objects  with Ks $<16$, the authors reached a very good precision on proper motions ($\sigma_\mu \sim$ 1 mas/yr) that is well-suited to discussing the internal dynamics of the cluster. Here we attempt to derive positions and proper motions linked to the International Celestial Reference Frame (ICRF, \cite{Ma(1998)}) for objects with Ks $<$ 18-19 by using for the reduction catalogues that are defined in this frame. This is at the price of a degraded precision because there are few reference stars, but it enables us to address the question of cluster membership including substellar objects down to M9 spectral type. 

This paper is organised as follows. In Sect. 2 we present the observational material used for the proper motion measurements. In Sect. 3 we present the reduction procedure and the resulting proper motion catalogue. In Sect. 4 we present the astrometric validation of the proper motions, in Sect. 5 the kinematic analysis of the $\rho$~Ophiuchi~F/E cores and in Sect. 6 a tentative classification of its content. We discuss the nature of the new members in Sect. 7. Our conclusions are summarised in Sect.8.


\section{Observational data}
The work that we present here aims at measuring proper motions in the near-IR inside an embedded cluster. These proper motions  are used to separate cluster members from background stars. We focused our efforts on the F/E cores of the Lynds L1688 cloud in the $\rho$~Ophiuchi star-forming region which is a region with high extinction (A$_{V}$ > 25-100 mag \citep{Wilking(2015)}) and ongoing star-formation.

The observations were carried out at the ESO-NTT equipped with the SOFI NIR camera. Astrometric (Ks) and photometric (J,H,Ks) observations were performed  in direct-imaging mode. A list of 15 pointing directions, in small- and large-field mode (SF, LF)  (see Table~\ref{centre}) was established to set up a mosaic covering the F and E cores with overlapping frames. The covered field is about 0.2$\degr$ x 0.2$\degr$. 

The observations were acquired during six observational epochs (2000.3, 2001.3, 2002.3, 2003.3, 2004.3, and 2009.3) with a total of 26 half-nights. The observations were performed at the same epoch of the year (April) so that no parallactic signature would perturb the astrometry.  We present the characteristics of the 648 SOFI exposures obtained for this work in  Table ~\ref{data}. Multiple exposures taken over three consecutive nights were performed at each epoch to average atmospheric effects and to enhance the signal-to-noise ratio (S/N).  

Two observing modes of SOFI were tried: nine in the SF mode, scale=0.144$\arcsec$/pixel, Field Of View (FOV) = 2.5$\arcmin$, and six in the LF mode, scale=0.288$\arcsec$/pixel, FOV=5$\arcmin$. Unfortunately, the SF observations appeared to be perturbed by distortions; the astrometry in the LF mode was much more stable. A table of distortions of the small field was set up from 30 frames taken in the direction of Omega Centauri, and the corrections were applied to our measurements. 

To complement our observational material we data-mined the public observational archives. We recollected a total of 789 additional NIR frames  (see Table~\ref{data}), obtained at various telescopes, that partially covered our field of work. These data allowed us to enlarge our field of work to about 0.5$\degr$ x 0.5$\degr$. 
Finally, we decided to also introduce the positions given by four global catalogues in our input data: USNO A2.0 \citep{USNO}, 2MASS \citep{2MASS}, WISE \citep{Wright(2010)}, and Spitzer (DR4) \citep{Gallagher(2003)}. Although the precision of these catalogues is poor in view of the present observational material, they enlarge the time base of our data and help to rigidify the local reference system. 

\begin{figure}[!htp]
\begin{center}
\includegraphics[width=0.49\textwidth]{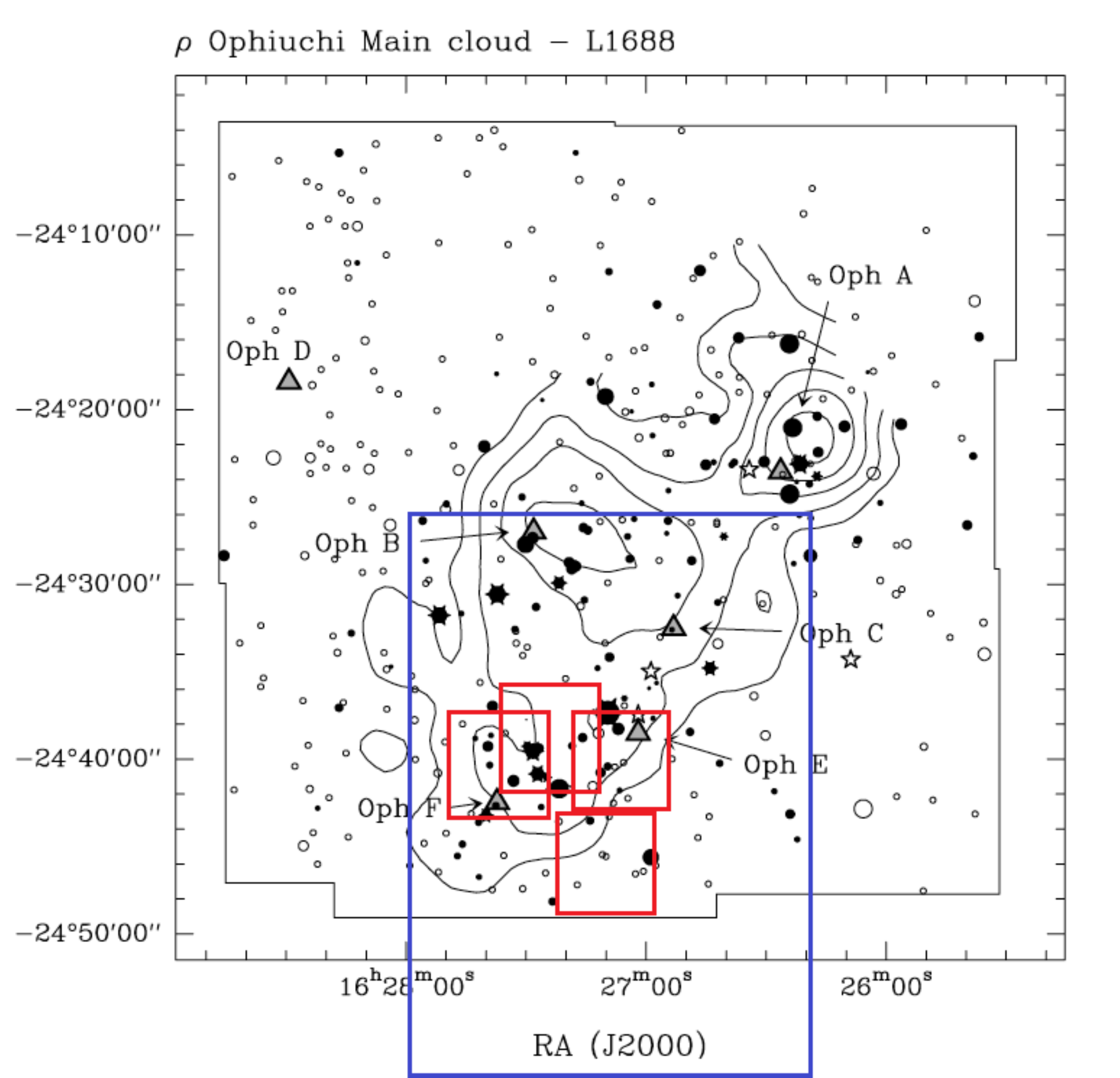}
\caption{\label{field} Areas covered by the present proper motion catalogue overlaid on the map of the L1688 dark cloud from \cite{Bontemps(2001)}. Red lines delimitate the NTT-SOFI observations and blue lines give the total extension of the catalogue including auxiliary data.} 
\end{center}
\end{figure}

We present in Fig. \ref{field} the field studied in this project overlapped on a map of the L1688 cloud from \cite{Bontemps(2001)}.

\begin{table}[!ht]
\caption{\label{centre}Pointing directions of the ESO-NTT(SOFI) observational program. }
\begin{tabular}{cccrc}
\hline
Right Ascension &Declination &SOFI\\
$\degr$      &    $\degr$              &     mode   \\
\hline
 246.77851	 & -24.74061 &SF	\\   
 246.78326	 & -24.68865 &SF	\\   
 246.78587	 & -24.71280 &SF	\\   
 246.79471	 & -24.65441 &SF	\\   
 246.79802	 & -24.69751 &SF	\\   
 246.81035	 & -24.69045 &SF	\\   
 246.85083	 & -24.69583 &SF	\\   
 246.88192	 & -24.66096 &SF	\\   
 246.91328	 & -24.66081 &SF	\\   
 246.78229	 & -24.72672 &LF	\\   
 246.79642	 & -24.75275 &LF	\\   
 246.80245	 & -24.67608 &LF	\\   
 246.83418	 & -24.68121 &LF	\\   
 246.85500     	 & -24.68139 &LF 	\\ 
 246.89746	 & -24.66093 &LF	\\   
       \hline
  \end{tabular}
\tablefoot{
SF and LF designate the small field and large field modes of the SOFI camera.}
\end{table}

\begin{table}[!ht]
\caption{\label{data}List of frames used to determine the proper motions. }
\begin{tabular}{cllrc}
\hline
Epoch&Telescope&Instr.&Nb&Scale\\
yr      &                 &        & frames   &$\arcsec$/pixel\\
\hline
  &     &      &    &   \\
2000 &NTT       &SOFI-SF   &  89 & 0.144 \\ 
2000 &NTT       &SOFI-LF  &   3 & 0.288 \\
2001 &NTT       &SOFI-SF   &  46 & 0.144 \\ 
2001 &NTT       &SOFI-LF  &  12 & 0.288 \\
2002 &NTT       &SOFI-SF   & 114 & 0.144 \\  
2003 &NTT       &SOFI-SF  & 148 & 0.144 \\
2003 &NTT       &SOFI-LF  &  12 & 0.288 \\
2004 &NTT       &SOFI-SF   & 103 & 0.144 \\
2004 &NTT       &SOFI-LF  &  27 & 0.288 \\
2009 &NTT       &SOFI-LF  & 148 & 0.288 \\
     \hline
\multicolumn{5}{c}{Auxiliary observations}\\
     \hline
  &     &      &    &   \\
2005 &UKIRT     &WFCAM     &  12 & 0.402 \\
2006 &UKIRT     &WFCAM     &   2 & 0.211 \\
2006 &UKIRT     &WFCAM     &   7 & 0.402 \\
2008 &AAT       &Iris2     &  11 & 0.447 \\
2004 &CFHT 3.6m &Megacam   &  15 & 0.187 \\
2006 &CFHT 3.6m &Megacam   & 103 & 0.187 \\
2007 &CFHT 3.6m &Megacam   &  21 & 0.187 \\
2011 &CFHT 3.6m &Megacam   &  13 & 0.187 \\
2007 &Subaru    &Suprimecam&  51 & 0.202 \\
2008 &KPNO 4.0m &NewFirm   &  79 & 0.395 \\ 
2009 &KPNO 4.0m &NewFirm   &  48 & 0.395 \\
2006 &CFHT      &WIRCam    & 425 & 0.303 \\
     \hline
\multicolumn{5}{c}{Catalogues}\\
     \hline
  &     &      &    &   \\
1982 &USNO A2.0 &	   &	 & \\
2000 &2MASS	&	   &	 & \\
2004 &Spitzer	&	   &	 & \\
2010 &WISE	&	   &	 & \\
     \hline
  \end{tabular}
\end{table}

%

\section{Data reduction}
All the frames we collected were measured using a dedicated source extraction code \citep{Viateau(1999)}  that has been optimised for astrometry purposes. As the absorption is very high in the region covered by the project, many images did not contain a sufficient number of stars to fit a Point Spread Function (PSF). Therefore we used a two dimensional Gaussian to extract the positions. 

We determined the proper motions through a global reduction of the whole data set collected for this project \citep{Ducourant(1991),Ducourant(2007),Teixeira(1992),Teixeira(1998)}. This means that we simultaneously derived the stellar parameters (position and proper motion) for the objects and the frame parameters of each of the 1437 frames. Beforehand the data were aligned on a common reference system with the help of an astrometric input reference catalogue.

The 2MASS catalogue \citep{2MASS} ($t_0$$\sim$2000) is generally a good option for this astrometric step since it is rather dense (about 11 000 sources per square degree on average on the entire sky) and its astrometric precision is acceptable (about 100 - 200 mas). Nevertheless, our project was spread over more than one decade and the 2MASS catalogue provides only positions around the epoch 2000 and no proper motions. It is therefore necessary to obtain proper motions estimates for this step, otherwise the dispersion of the proper motions after reduction is so large that no membership analysis can be performed.

\subsection{Input astrometric reference catalogue}
 We then decided to set up an input astrometric reference catalogue to align the plate measurements onto a common reference frame. We used the 2MASS positions collected in the work zone that we complemented with proper motions derived as follows. We first searched in the PPMXL \citep{Roeser(2010)} and SPM4 \citep{Girard(2011)} for visible counterparts of the 2MASS objects and took the proper motion with the best internal precision when any was available. In this way, we were able to identify 259 objects. These objects typically lie at the periphery of our field where the absorption due to dust is much lower and objects are detectable in the visible wavelengths. For the vast majority of objects, which are only detectable in the infrared or NIR, we combined the positions from 2MASS with those from the AllWISE data release \citep{Cutri(2013)} (central epoch$\sim$2010), matched in a 3$\arcsec$ search radius to estimate the proper motions. We excluded the AllWISE objects with a semi-major axis of the error ellipse larger than 200 mas, however. We derived more than six hundred proper motions estimates in the NIR in this way.

The resulting input astrometric catalogue contained positions and proper motions for about nine hundreds objects and was used to scale and rotate the 1437 frames and align them onto the axes of the input catalogue. The rescaled and rotated lists of measurements were then cross-correlated and compiled in a metalist that contained the measurements of each object in each image in which it was detected. This metalist contains 2205 objects.

\subsection{Proper motions}
Then the whole set of measurements of the objects detected in the $\rho$ Ophiuchi region was globally reduced through a block-adjustment-type iterative procedure described in \cite{Ducourant(2007),Ducourant(2008)}, which allowed us to simultaneously compute the unknown parameters of all stars (positions and proper motions) and the unknown plate parameters of all frames.

When we examined the residuals of the global fit as a function of the  observations epochs, we observed that the residuals appeared to systematically deviate from the expected null mean value at three epochs. One deviation in 1982 corresponds to the USNO A2.0 positions (see \cite{Assafin(2001)} for an analysis of systematics of the USNO A2.0), another in 2004 corresponds to Spitzer positions and the last one in 2010 to the WISE positions. These deviations  indicate a local systematic effect in these catalogues. We therefore applied the following corrections to these data, which correspond to the mean of the residuals :  $(\Delta\alpha,\Delta\delta)$= (-182,+86) mas for USNO A2.0 positions, (+152,-217) mas for Spitzer positions, and (+23,-137) mas for WISE positions. After applying these corrections, the  input reference catalogue that contains the proper motions based on WISE and 2MASS positions was regenerated using the modified positions, and the whole data set was reprocessed.

An a posteriori analysis of the derived proper motions showed evidence that the quality of the plate-to-plate transformations was degraded in the regions where cluster members were too numerous with respect to field stars. We therefore decided to exclude objects with a much higher velocity than the mean velocity of field stars (as defined in Sect. \ref{membership}) from the plate-to-plate transformations because these objects are probably cluster members or foreground stars. 

\begin{figure}[!htp]
\begin{center}
\includegraphics[width=0.49\textwidth]{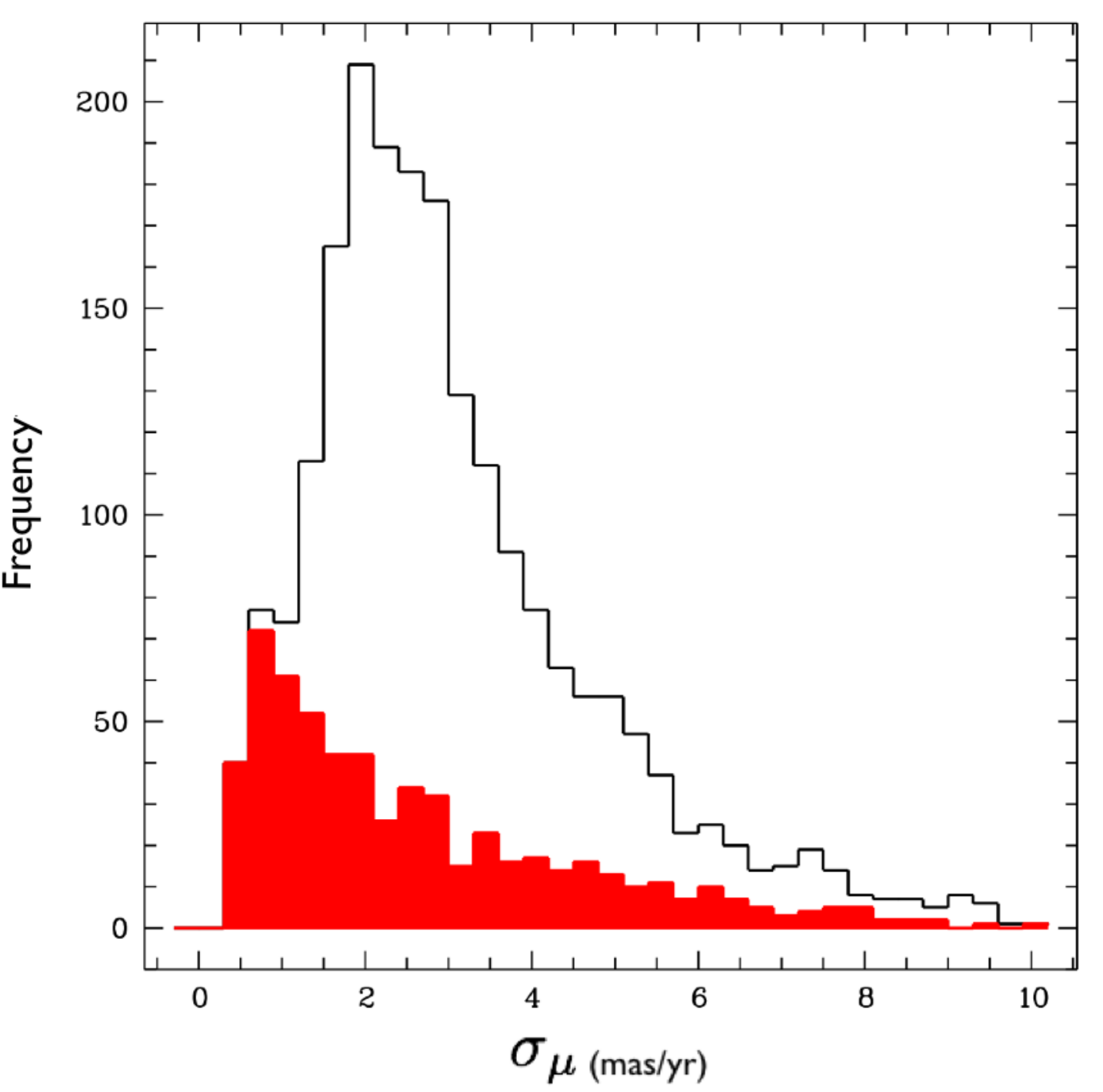}	
\caption{\label{histsmu} Histogram of internal precisions of our proper motion catalogue in the $\rho$~Ophiuchi embedded cluster. The solid black line corresponds to the whole catalogue while the red filled histogram designates objects which benefit from SOFI data (SOFI catalogue).} 
\end{center}
\end{figure}

We finally obtained positions and proper motions for 2213 objects spread over a 0.5$\degr$ x 0.5$\degr$ field. These objects are separated into two groups :  607 are located in a central region of 13.8$\arcmin$ x 10.8$\arcmin$ and benefited from SOFI observations, while the remaining objects have proper motions derived only from auxiliary data. We present the distribution of the internal precisions of our proper motions in Fig.~\ref{histsmu}. 

The spread of the precisions is the reflection of the historical material proper to each object and of its magnitude. It is obvious from this figure that the proper motions derived with the contribution of SOFI observations (red filled histogram) are more precise than the ones obtained from auxiliary data. The subsample benefiting of the SOFI data will constitute the core of our catalogue that we will designate hereafter as the  SOFI catalogue. 

\subsection{Photometry}
\begin{figure}[!htp]
\begin{center}
\includegraphics[width=0.49\textwidth]{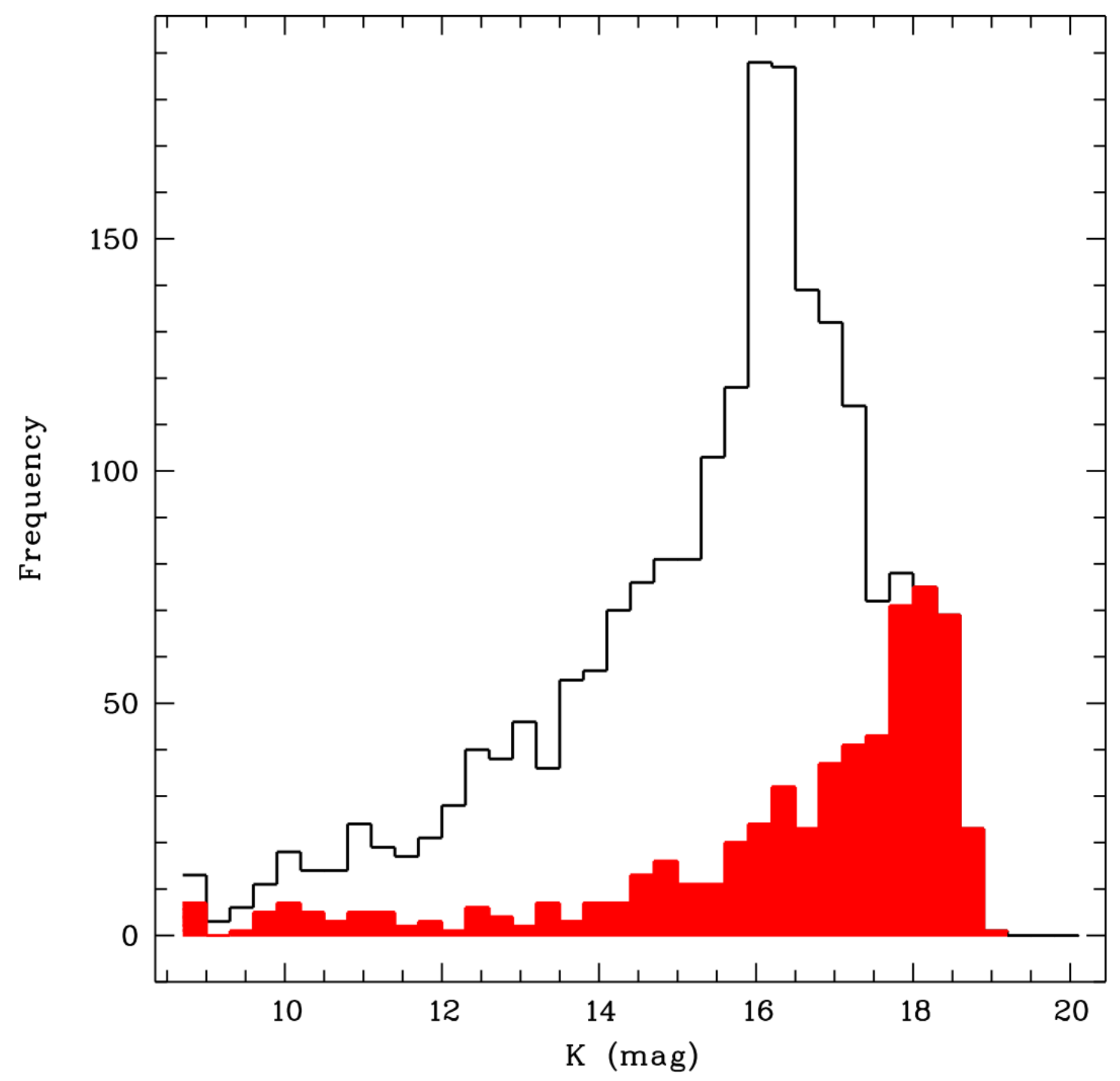}	
\caption{\label{histK} Histogram of K$_s$ magnitudes of the proper motion catalogue. The black line correspond to the whole catalogue while the red filled histogram designates the SOFI catalogue.} 
\end{center}
\end{figure}

LF SOFI observations were acquired in the J, H, Ks bands (central=(1.247,16.653,2.162) $\mu$m, width=(0.290,0.297,0.275) $\mu$m) to complement the 2MASS photometry for fainter objects of the SOFI catalogue. We present the distribution of the Ks magnitudes of objects in our catalogue in Fig.~\ref{histK} and list in Table \ref{photo} the J,H,Ks photometry from 2MASS or derived in this work for the kinematic members of the $\rho$~Ophiuchi core (see next section for the membership determination) together with the AllWISE \citep{Cutri(2013)} (w1,w2,w3,w4) photometry. 

The  SOFI catalogue explores the cluster down to much fainter magnitudes than the remaining catalogue and reaches K$_s$=18.6 mag. 

\begin{figure}[!htp]
\begin{center}
\includegraphics[width=0.49\textwidth]{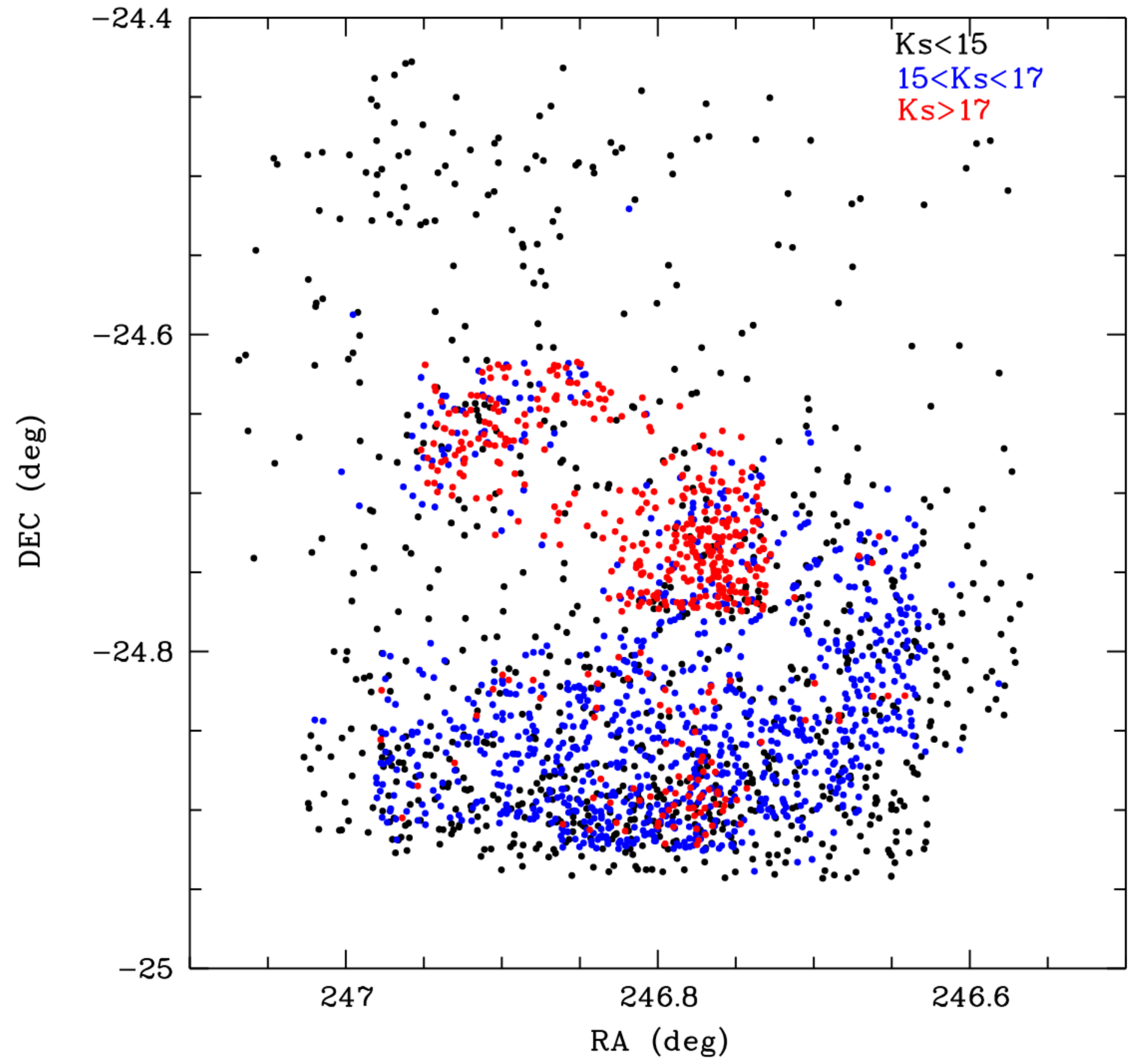}	
\caption{\label{carte-K} Distribution on the sky of objects from our proper motion catalogue with indication of their magnitude range.} 
\end{center}
\end{figure}

It is also interesting to note that the distribution on the sky of the various classes of magnitudes is not homogeneous, as shown in Fig. \ref{carte-K}. The faintest objects are essentially detected where the deeper SOFI observations are present and coincide with the cluster region where the extinction is high, while the brighter objects essentially lie at the periphery of the cluster. 

\section{Astrometric validation}
It  is difficult to asses the external errors of our catalogue because two thirds of our catalogue concern objects that are invisible at optical wavelengths and because to our knowledge, no astrometric proper motion reference catalogue exists in the NIR in this region. The recent release of the AllWISE catalogue \citep{Cutri(2013)} provides proper motion estimates in the NIR. Unfortunately, their quality is far too poor to be helpful in a region such as the $\rho$ Ophiuchi complex (the mean standard error on these proper motions in the studied region is is $\sigma_{\mu}\sim$300 mas/yr).

\subsection{Comparison to PPMXL and 2MASS catalogues}	
For the optical part of our catalogue, the densest comparison catalogue is the PPMXL catalogue \citep{Roeser(2010)}. We note that the present work included some PPMXL proper motions as starting point of the iterative reduction process but the final catalogue should be more or less independent of it. Another difficulty results from the fact that the 239 objects common to our catalogue and PPMXL are mostly concentrated at the southern edge of the field, which means that the cluster is almost completely excluded from this comparison and only the brightest objects of our catalogue are concerned. Nevertheless, we performed the comparison of both sets of proper motions, and we noted a systematic mean difference (in the sense \textit{this work} minus PPMXL) of ($\Delta\mu_{\alpha}cos(\delta)$,$\Delta\mu_{\delta}$)=(-1,+6) mas/yr. 

\begin{figure}[!htp]
\begin{center}
\includegraphics[width=0.45\textwidth]{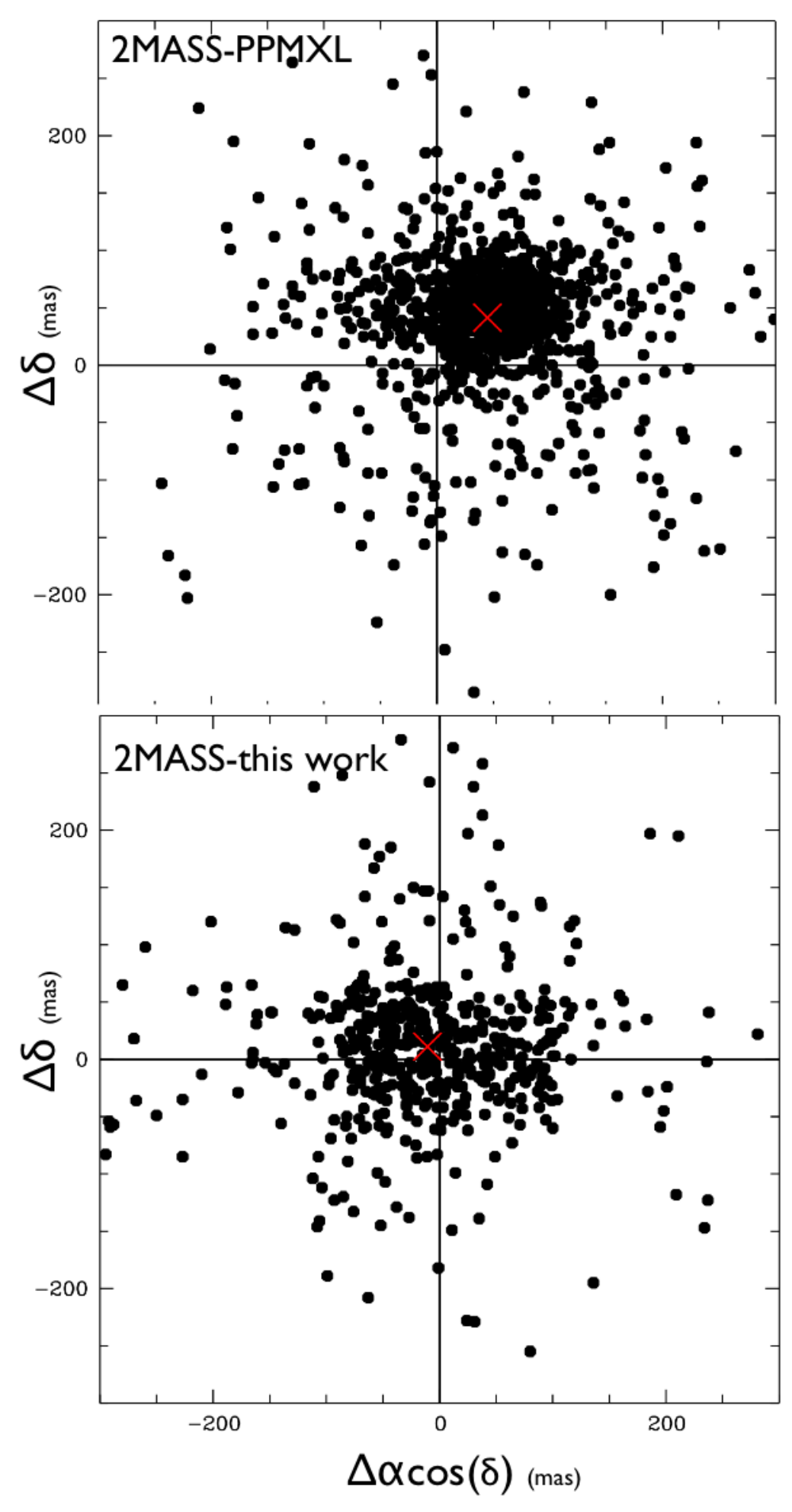}	
\caption{\label{comp2} Comparison of 2MASS positions with PPMXL positions (upper panel) and with this work (lower panel) at 2MASS mean epoch ($\sim$epoch 1999.3).  Red crosses indicate the means of the distributions after a 3$\sigma$ elimination.} 
\end{center}
\end{figure}

To determine the origin of the large systematic difference (essentially in declination), we considered the 2MASS catalogue which provides positions for epoch $\sim$1999.3 with errors $\sim$100 mas and compared its positions to the PPMXL and to our catalogue, each transported to the 2MASS epoch by application of its own proper motions. We present these comparisons in Fig. \ref{comp2}. The resulting mean differences in position in the sense 2MASS minus \textit{catalogue} are (+45,+41) mas for PPMXL (1321 objects) and (-11,+11) mas (504 objects) for this work. 

When we assume that 2MASS provides positions without bias, we can conclude that both catalogues suffer from systematic effects in their proper motions :  $\sim$(+2.7,+2.5) mas/yr for PPMXL (mean epoch $\sim$1982.6) and $\sim$(+1.3,-1.3) mas/yr for this work (mean epoch $\sim$2007) and that a large part of the differences observed in the comparison of our proper motions with PPMXL is due to PPMXL.

We note that the PPMXL and the present catalogue incorporated the 2MASS catalogue at some stage of their elaboration, which diminishes the impact of this comparison. Moreover, the comparisons presented in this section (PPMXL/\textit{this work}, PPMXL/2MASS, \textit{this work}/2MASS) do not rely on the same set of objects, and this may explain the larger systematic difference observed in the comparison of the PPMXL/\textit{this work}.

\subsection{Comparison to the Besan\c con Galaxy model}
Another way to investigate the possible systematic effects contained in our catalogue is to compare its mean behaviour with that of theoretical predictions given by the Besan\c con Galaxy model simulations \citep{Robin(2003),Robin(2004)} (this model does not include any cluster). For this purpose, we selected the set of 787 objects in our catalogue that are classified as field members in our membership analysis with a probability greater than 90 $\%$ (see Sect. \ref{membership} for the detail of filtering and membership probability calculation). When we compare the mean proper motion of these stars to the Besan\c con simulation (Ks<18.5 mag, d<50 Kpc) we observe mean differences of about 0.5 mas/yr on both coordinates, which is coherent with the previous analysis of systematic effects. 

\subsection{Analysis of the kinematic properties of the cluster members} 
Recently, \cite{Wilking(2015)} published a very precise catalogue of \textit{relative} proper motions in the $\rho$ Ophiuchi cluster. They provided relative proper motions for field objects and for known cluster members and evaluated their median uncertainties to 0.60 mas yr$^{-1}$ in RA and 0.71 mas yr$^{-1}$ by analysing the dispersion of the proper motions of the known cluster members (the expected velocity dispersion in the cluster is about 1 km/s i.e., $\sim$ 1 mas/yr). 

We applied the same method here and analysed the proper motion mean properties of the 68 cluster members (see Sect.\ref{membership}) of our catalogue with the highest membership probabilities (Prob>90$\%$). The dispersion in proper motion is ($\sigma \mu_{\alpha} cos(\delta)$,  $\sigma \mu_{\delta}$) = (4.7,3.8) mas/yr. This result is representative of the external precision of the present proper motion catalogue. After various investigations we concluded that the main limitation to the precision of our proper motions results from the inhomogeneous spatial distribution of stars and the low quality of the input astrometric catalogue used to set up the scale and the orientation of the frames. 

We also analysed the mean proper motion of field stars (probability>90$\%$) (see Sect. \ref{membership}) in the various SOFI fields. We could observe that the mean properties of these stars varied from one region to the other by 1-2 mas/yr, indicating possible local systematic effects in our catalogue.

\subsection{Conclusion on accuracy and systematic effects }
From these various tests we can conclude that our catalogue has global systematic effects of $\sim$2 mas/yr. The spatial repartition of stars is uneven in the field and was the source of the problem during the plate-to-plate connection, essentially in the cluster zone, which generated local systematic effects in the proper motions of about 1-2 mas/yr, and degraded the accuracy. Moreover, the stars in the cluster zone are essentially faint, which complicates reducing the images even more; the resulting accuracy of these proper motions is therefore poorer than the internal errors might suggest. 
\section{Kinematic analysis}
\subsection{Membership analysis}\label{membership}
To separate cluster members from field stars, we performed a membership analysis based on the results reported by \cite{Cabrera(1985)}. This paper, derived from \cite{Sanders(1971)}, itself inspired by the method of  \cite{Vasilevskis(1958)} allows establishing membership probabilities in open clusters from proper motion data.  The model consists of a mixture of two normal bivariate distributions, a circular one for the cluster and an elliptic one for the field stars. The selection of members and the derivation of the kinematic properties of the two distributions are performed simultaneously through an iterative process and a 50\% membership probability criterion.  Hereafter we distinguish  \textit{members} that have a high probability ($P$>0.9) from \textit{candidate members} (0.5<$P$<0.9). 

In the present work, the quality of proper motions is heterogeneous and it seems reasonable to filter the best measured objects to derive the kinematic properties of the cluster and of the field. On the one hand, it may also be interesting to test any object of the catalogue for its membership to the F/E cores although its proper motion is not of high accuracy to have a census of members as complete as possible. This is why we performed our analysis in two steps. 

We first selected the most reliable proper motions of our SOFI catalogue to derive the kinematic properties of the cluster and of the field, using objects with at least a time base of observations of five years, with a number of observations greater than 20, and with $\sigma_{\mu} \le 2$mas/yr. From these, we excluded proper motions higher than 45 mas/yr, which correspond to foreground objects that do not belong to the $\rho$ Ophiuchi complex. This left us with 342 objects with highly accurate proper motions. 

The membership analysis led to the following mean proper motions and precisions for the cluster derived with 48 members ($\mu_{\alpha}^*$, $\mu_{\delta}$)$_{cl}$= (-7.4 -22.9) $\pm$0.8 mas/yr (hereafter $\mu_{\alpha}^*$ stands for $\mu_{\alpha}cos\delta$) and for the field  ($\mu_{\alpha}^*$, $\mu_{\delta}$)$_{f}$ = (-3.5, -1.4) $\pm$ (0.4,0.3) mas/yr. 
The kinematic values derived for field stars agree reasonably well with predictions from the Besan\c con Galaxy model simulations in the direction of our field~:~($\mu_{\alpha}^*$, $\mu_{\delta}$)= (-2,-2) mas/yr.

In a second step, we considered all the objects from our proper motion catalogue that were excluded from previous analysis and selected those whose data were spread over five or more observational epochs with a time base of observations longer than six years and a precision better than 3.5 mas/yr. We then tested their membership to the two distributions and retained those with a probability greater than 50$\%$. 

\begin{figure}[!htp]
\begin{center}
\includegraphics[width=0.49\textwidth]{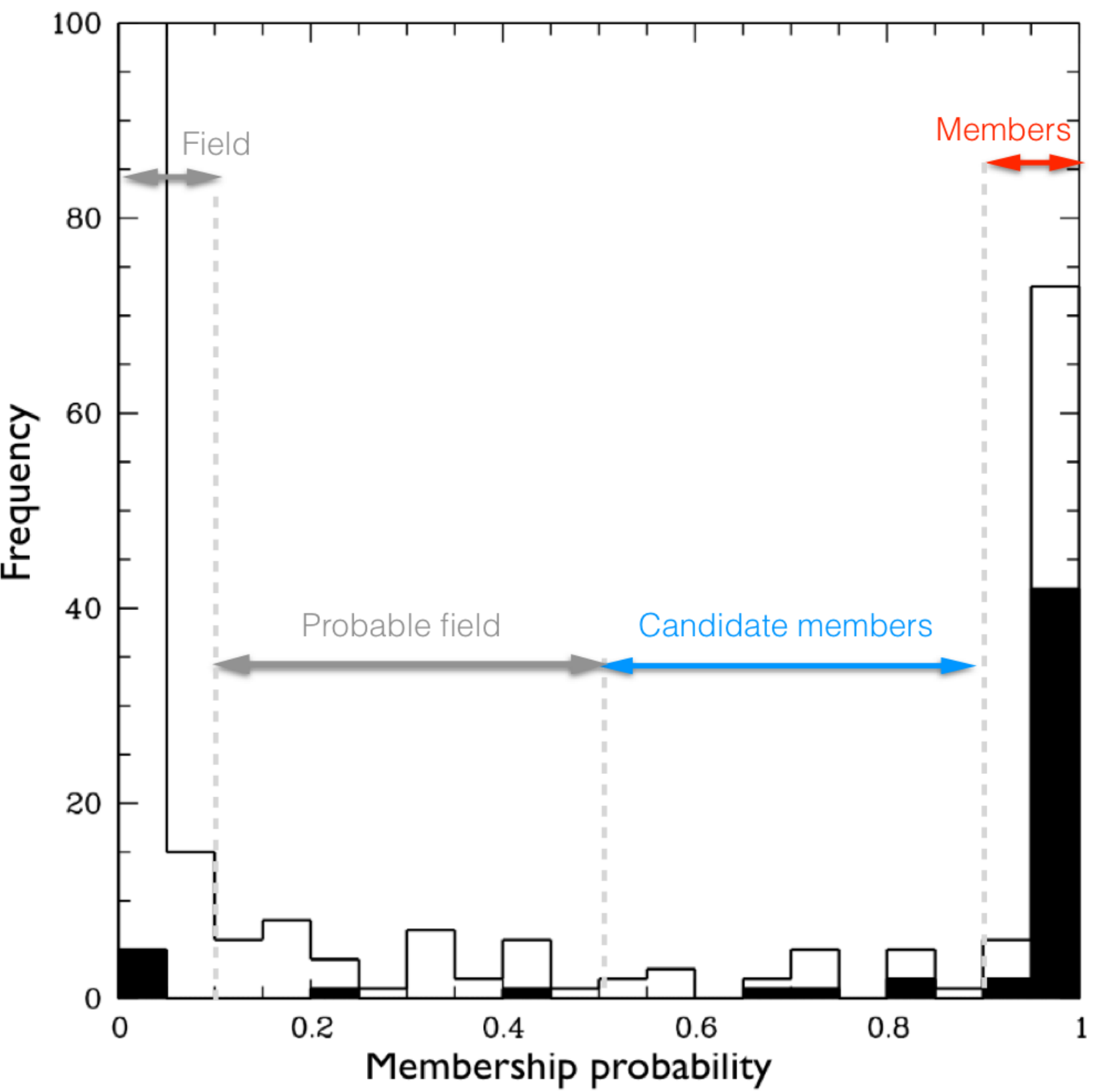}	
\caption{\label{histprob} Repartition of membership probabilities of the 936 objects.  The filled black histogram corresponds to known young stellar objects (detected by ISO \citep{Bontemps(2001)} or Spitzer \citep{Evans(2003)}) that are present in our catalogue. For visibility the y axis is truncated; the peak that corresponds to a probability of $\sim$0.05 culminates at 800 objects.}
\end{center}
\end{figure}

We thus end up with 82 objects with a membership probability to the cluster higher than 50\%. We present the repartition of membership probabilities for the 936 objects analysed here in Fig.\ref{histprob}. We observe in this figure two clear peaks that correspond to field population ($P$<0.1) and cluster population ($P$>0.9) and objects with intermediate probabilities. Sixty-eight objects have a probability P>0.9 and are considered as \textit{members}, while14 objects have a membership probability 0.5<$P$<0.9 and are classified as \textit{candidate members}. 

Membership analyses are very sensitive to outliers (e.g., foreground objects from the Galaxy with high proper motion) and need to be excluded to derive realistic solutions (see discussion about the pruning of the data in \cite{Cabrera(1985)}). We therefore excluded objects with $\mu_{\alpha}^*$ or $\mu_{\delta}$ > 45 mas/yr at step one and objects with proper motions beyond 2.5 $\sigma$ of the mean properties of the cluster and of field at the end of step two. We present a list of objects in Table \ref{elim} that are considered as outliers despite their high membership probability ($P$>0.9) because their proper motion differs by more than 2.5$\sigma$ from that of the cluster.

\begin{figure}[!htp]
\begin{center}
\includegraphics[width=0.49\textwidth]{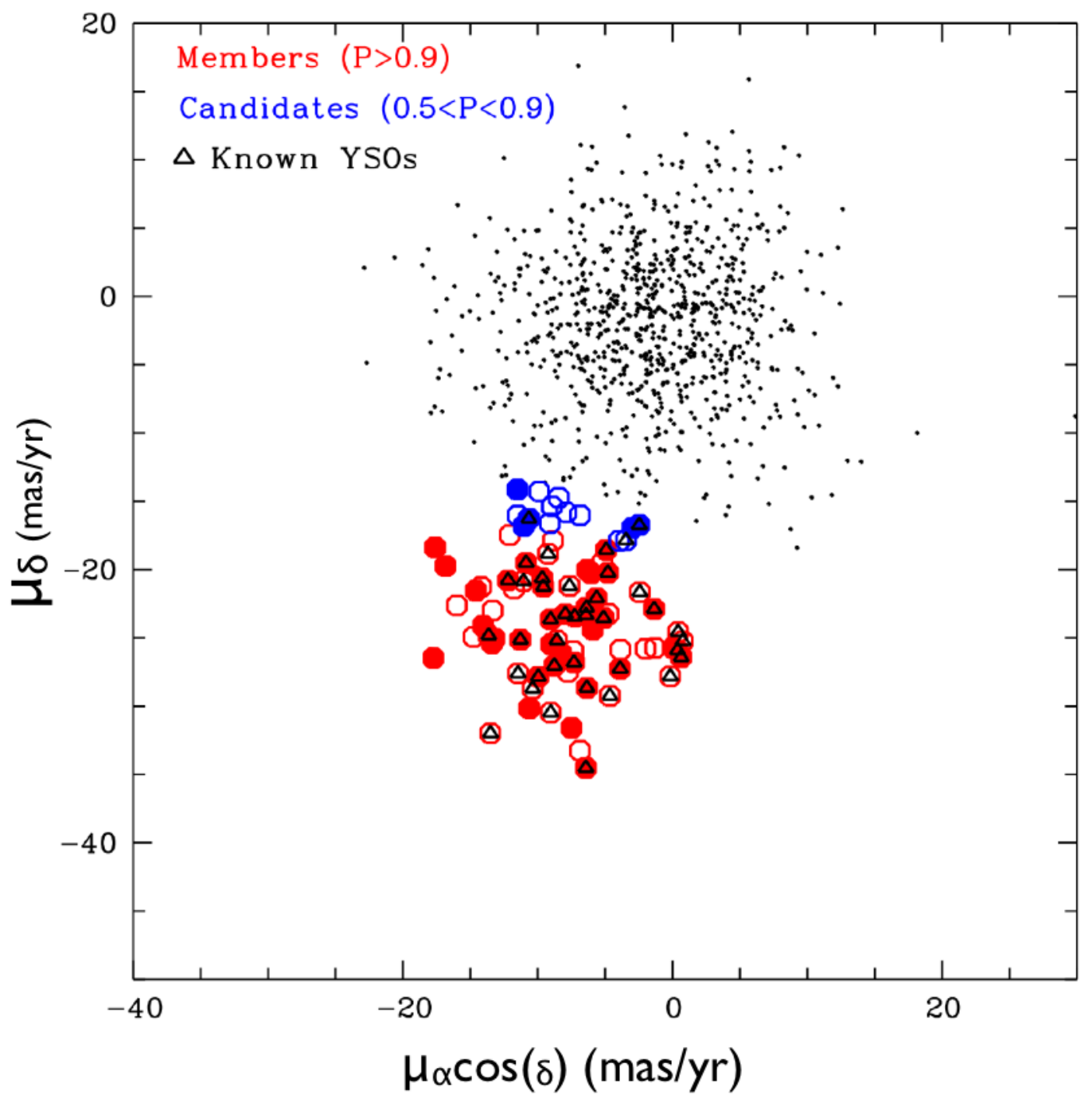}	
\caption{\label{vpd} Vector plot diagram. Red circles correspond to the kinematic \textit{members} (P>0.9) derived in this work, blue circles to \textit{candidate members} (0.5<P <0.9). The black dots correspond to field objects (P<0.1). Filled circles designate objects belonging to the SOFI catalogue and triangles identify known young stellar objects detected by ISO or Spitzer.}
\end{center}
\end{figure}

\begin{figure}[!htp]    
\begin{center}
\includegraphics[width=0.49\textwidth]{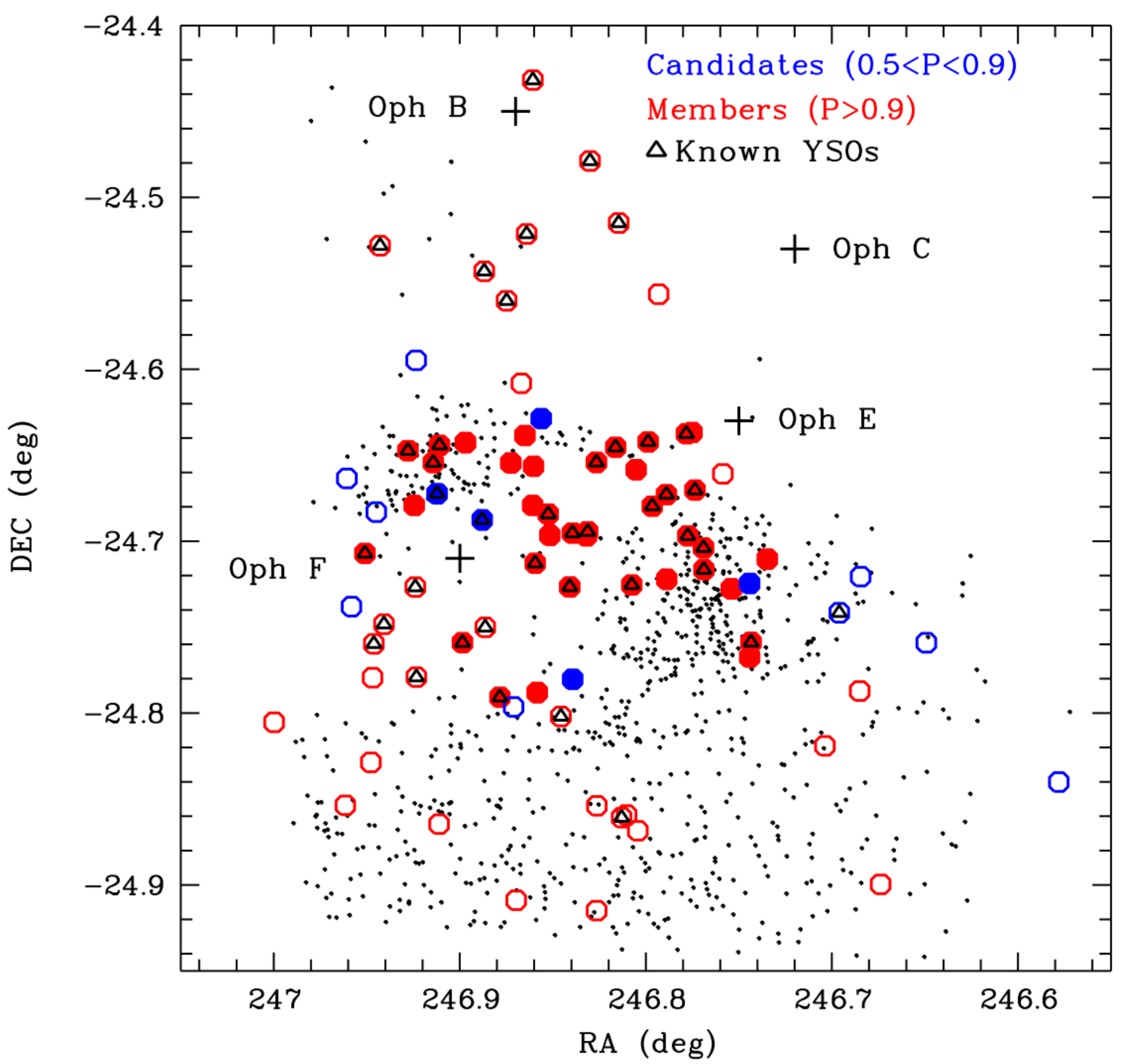}	
\caption{\label{radecprob} Distribution on the sky of kinematic \textit{members}, \textit{candidate members} (0.5<P<0.9) and field stars with indication of the localisation of the B, C, E, and F cores of $\rho$ Ophiuchi. Red circles correspond to the kinematic \textit{members} (P>0.9) derived in this work, blue circles to \textit{candidate members} (0.5<P <0.9). Filled circles designate objects belonging to the SOFI catalogue and triangles identify known young stellar objects detected by ISO or Spitzer. The black dots correspond to field objects (P<0.1). }
\end{center}
\end{figure}

The resulting mean kinematic properties of the cluster are listed in Table \ref{properties}. We present the list of the \textit{members} and \textit{candidate members} with their astrometric solution and their membership probability in Table~\ref{members}. Fig.~\ref{vpd} shows the vector plot diagram of the cluster and indicates the membership probability, and Fig.~\ref{radecprob} shows the distribution of field stars and of cluster \textit{members} and \textit{candidate members} across our field.

Figure \ref{vpd} shows that the repartition of the kinematic \textit{members} is globally circular, as expected. The dispersion of their proper motions ($<5$ mas/yr) is larger than expected from the astrophysical point of view (1-2 mas/yr) and reflects the difficulties of measuring proper motions in such obscured zones and the reality of our accuracy. Twenty-six members are YSOs that have been identified by ISO or Spitzer. Field stars and cluster \textit{members} are separated in Fig.\ref{vpd} by a layer of objects with intermediate membership probabilities (\textit{candidates}).

\subsection{New and known members}
Of the 82 kinematic \textit{members} or \textit{candidate members} derived in this work, 56 are referenced at CDS as young stellar objects, the 26 others are new \textit{members} or new \textit{candidates}. Of the known objects, 7 are brown dwarfs (BDs) \citep{Alves(2012)}, 1 is a BD candidate  (FHTWIR-Oph 80, \cite{Alves(2010)}), and 26  were found by the major spatial NIR-surveys ISO \citep{Bontemps(2001)} or Spitzer \citep{Evans(2003)}. 

The kinematic members are mostly located around the Oph F and E cores, with several new detections south to these cores. Some previously known members around Oph B core are also classified as kinematic members. The high concentration of kinematic members in the central part of the field covered by this study corresponds to the zone of deep SOFI observations. 

\begin{figure}[!htp]    
\begin{center}
\includegraphics[width=0.40\textwidth]{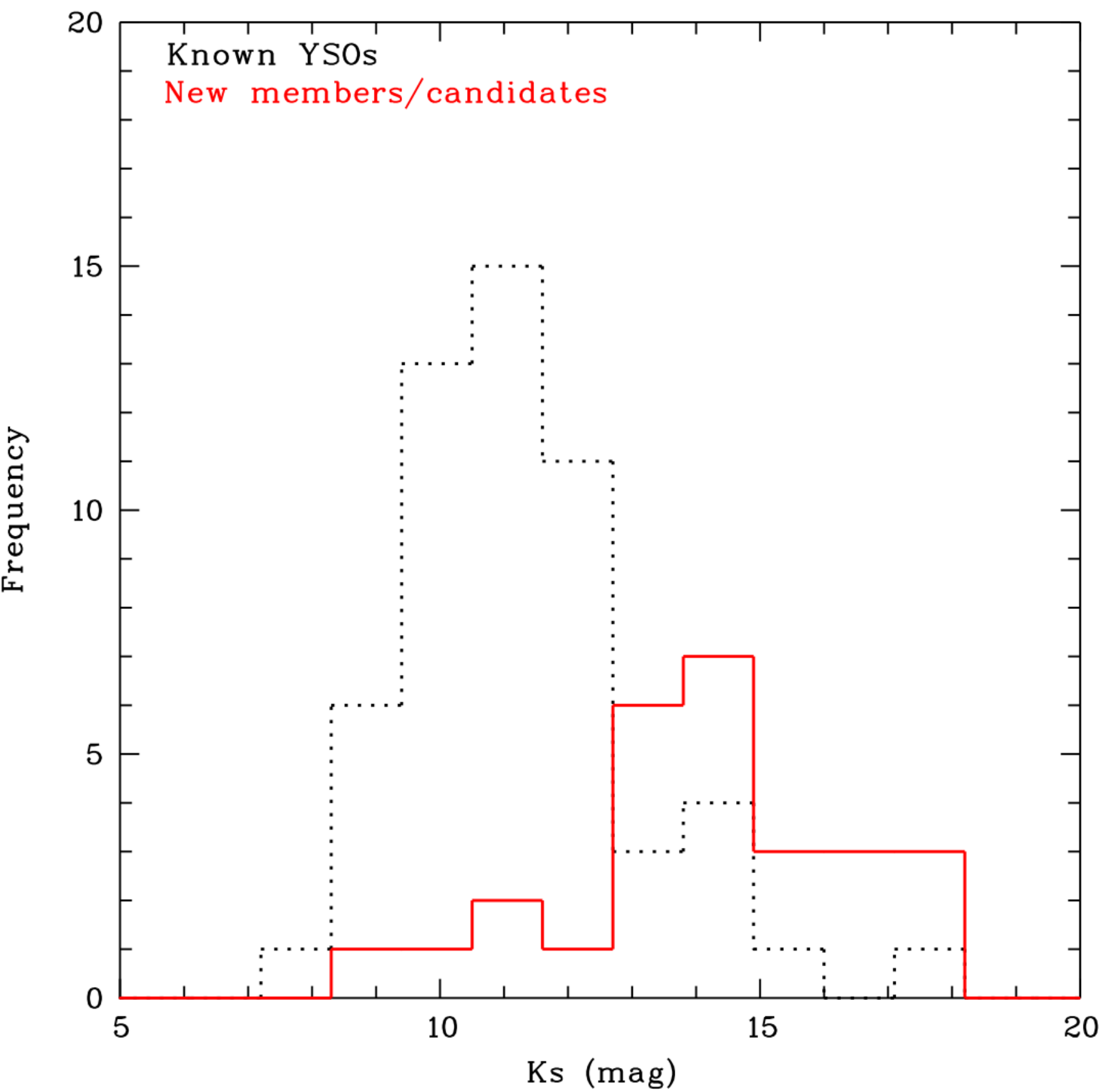}	
\caption{\label{status} K$_s$ distribution of known young stellar objects referenced at the CDS and new kinematic members or candidates of the cluster.}
\end{center}
\end{figure}

We present in Fig. \ref{status} the repartition in terms of K$_s$ magnitudes of the 56 known and 26 new members.

The new members appear essentially fainter (down to Ks=18.1)  than the known members and were revealed by the deep Near-IR observations performed in this work.

We note that some well-attributed members of $\rho$ Ophiuchi are not included in our list of kinematic members because their astrometry was too poor to analyse them.

\subsection{Spatial velocities}
We derived the Galactic coordinates (X towards the Galactic centre, Y in direction of Galactic rotation, Z towards the North Galactic Pole) and heliocentric space velocity of the cluster using  the most recent distance determination d=133.3$\pm$3.6 pc \citep{Ortiz(2015)}  and the median radial velocity for Oph pre-main-sequence stars (-6.3$\pm$0.3) km s$^{-1}$ \citep{Prato(2007), Guenther(2007), Kurosawa(2007), James(2006)}. We summarise in Table \ref{properties} the kinematic properties of the $\rho$ Ophiuchi (F core). 

\begin{table}[!ht]
\caption{\label{properties}Mean spatial and kinematic properties of the $\rho$ Ophiuchi F core as derived from the 68 kinematic members with the highest membership probability (P>0.9). }
\begin{tabular}{ll}
\hline
Property 										&Value			      			\\
\hline
($\alpha$,$\delta$)								 & ( $16^h$ $27^m$ $22.51^s$, $-24\degr$ $41^`$ $58.5^"$)	\\
($ \mu_{\alpha}$ cos$\delta$, $\mu_{\delta} $) 		         &  ( -8.2, -24.3)$\pm$ 0.8 mas/yr 		\\
(l,b)											& ( 352.9816\degr,16.5540\degr)	\\
($ \mu_{l}$ cos b, $\mu_{b}$)						&  ( -23.8,-10.7)$\pm$ 0.8 mas/yr 		\\
Position $(X,Y,Z)$								& ( 126.9, -15.6, +37.4) pc				\\
Velocity $(U,V,W)$								& ( -5.9$\pm$0.1, -14.2$\pm$0.3, -8.1$\pm$0.4) km/s\\
 \hline
  \end{tabular}
\tablefoot{The mean proper motion of the core given here does not account for the probable systematic error ($\sim$ 2 mas/yr) of our proper motion catalogue.}
\end{table}

We compared these values with the kinematic characteristics of the Upper Scorpius (US) association since the $\rho$ Oph cluster is expected to be part of the Upper Sco association (see \cite{Wilking(2008)}). We observe that the velocity derived here for the cluster is very similar to the one that we derive for the sample of 117 members of Upper Sco OB2 group with Hipparcos measurements \citep{deZeeuw(1999)} (U,V,W)$_{US}$ = (-5.0 , -14.8 , -6.5 ) $\pm$ (0.1, 0.4, 0.2) km/s. 

\section{Classification}

Of the 82 kinematic \textit{members} or \textit{candidate} members,  29 have been classified in terms of their spectral energy distribution (SED) into class I, II, or III (\cite{Bontemps(2001)} and \cite{Gutermuth(2009)}), but 53 of them appear to be unclassified. Some of these unclassified members are referenced as YSOs at the CDS. 

Accordingly, we decided to perform a tentative classification into the three most significant YSO classes (I, II and III) using photometric data from Table \ref{photo}.  If there are some contaminants in our kinematic members sample, then they will be incorrectly attributed to one of the three YSO classes since we assume here that each of the 82 members and candidates are young stellar objects. 

For the classification we adopted a supervised learning method, whose principle is the following. The method considers the various magnitudes of each previously classified member (J,H,K$_s$,w1,w2,w3,w4) and all their colour combinations (J-H, J-K$_s$, J-w1, etc.) and searches for regions in this magnitude-colour hyperspace that are occupied by members of a same class (I, II and III). When the parameters characterising these regions are ``learned', the algorithm considers the unclassified members and attributes a class to them depending on their localisation in this hyperspace.

Of the various supervised methods available, we adopted the well-known and successful data classification non-parametric method called random forests \citep{Breiman(2001),Breiman(2003)}, complemented by a Monte Carlo trial to account for errors in the observable quantities. We adopted the R language implementation of random forests by \cite{Liaw(2002)}, and to take all the errors of the available measurements into account, we followed a Monte Carlo approach, performing 30000 independent classifications of the dataset. At each run the following process was performed: first the magnitudes of each star at each filter are sampled from Gaussian probability distribution functions with the mean equal to the measured (or imputed, see the paragraph below) value and the sigma equal to the measured (or imputed) magnitude error at the relevant filter; then all the possible colour combinations between the filters J, H, K, w1, w2, w3 and w4 are computed; then a random forests classifier assigns a class to all the objects with unknown classes. Finally, after all the independent runs, the frequency that each object was assigned to each class is computed and the class with the greatest frequency is adopted as the object class.

A special approach had to be adopted because some data points were missing in our dataset (e.g., many objects were observed or detected only in some, but not all, of the 2MASS or WISE filters). The approach adopted was to perform data imputation and Monte Carlo sampling of error distributions. The first step is to perform data imputation, but to avoid mixing the well-known objects with the most uncertain objects, we split the missing data imputation into two phases: first, the data imputation was performed using all the objects with previously known classifications; then, the missing data imputation was performed considering all objects in our dataset. We adopted a nonparametric missing data imputation method called missForest \citep{Stekhoven(2012)}. This method is also based on a random forests algorithm to predict continuous missing values from the observed values of all other objects in the dataset and from the relation of the object whose value is missing and all other objects considering the entire parameter space of the dataset.

\begin{figure}[!htp]    
\begin{center}
\includegraphics[width=0.49\textwidth]{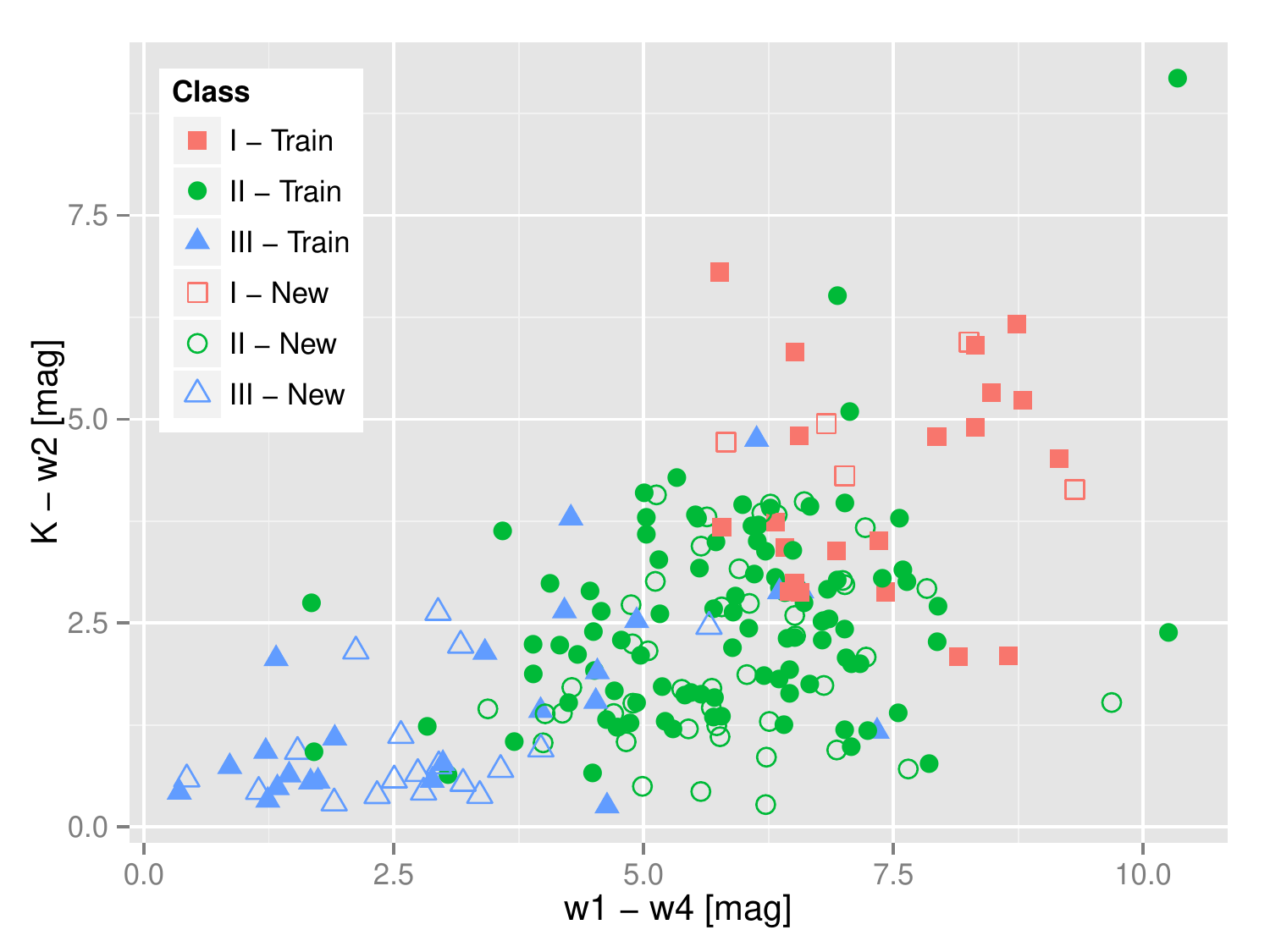}	
\caption{\label{colour}One of the colour - colour diagrams for kinematic members using 2MASS K$_s$ and WISE w1,w2, and w4 photometry. Objects with classes previously known in the literature that were used for the training of the method (train) are shown with filled symbols and the objects we classified are shown with empty symbols. Squares refer to class I, circles to class II, and triangles to class III.}
\end{center}
\end{figure}

The results of the classification performed here are represented in one of the many possible colour-colour planes in Fig. \ref{colour} which shows that most of the objects lacking literature classification were assigned a class II status. Of the 53 unclassified \textit{members} or \textit{candidates}, 16  were assigned a class III status, 32 a class II, and 5 a class I. 

\begin{figure}[!htp]    
\begin{center}
\includegraphics[width=0.49\textwidth]{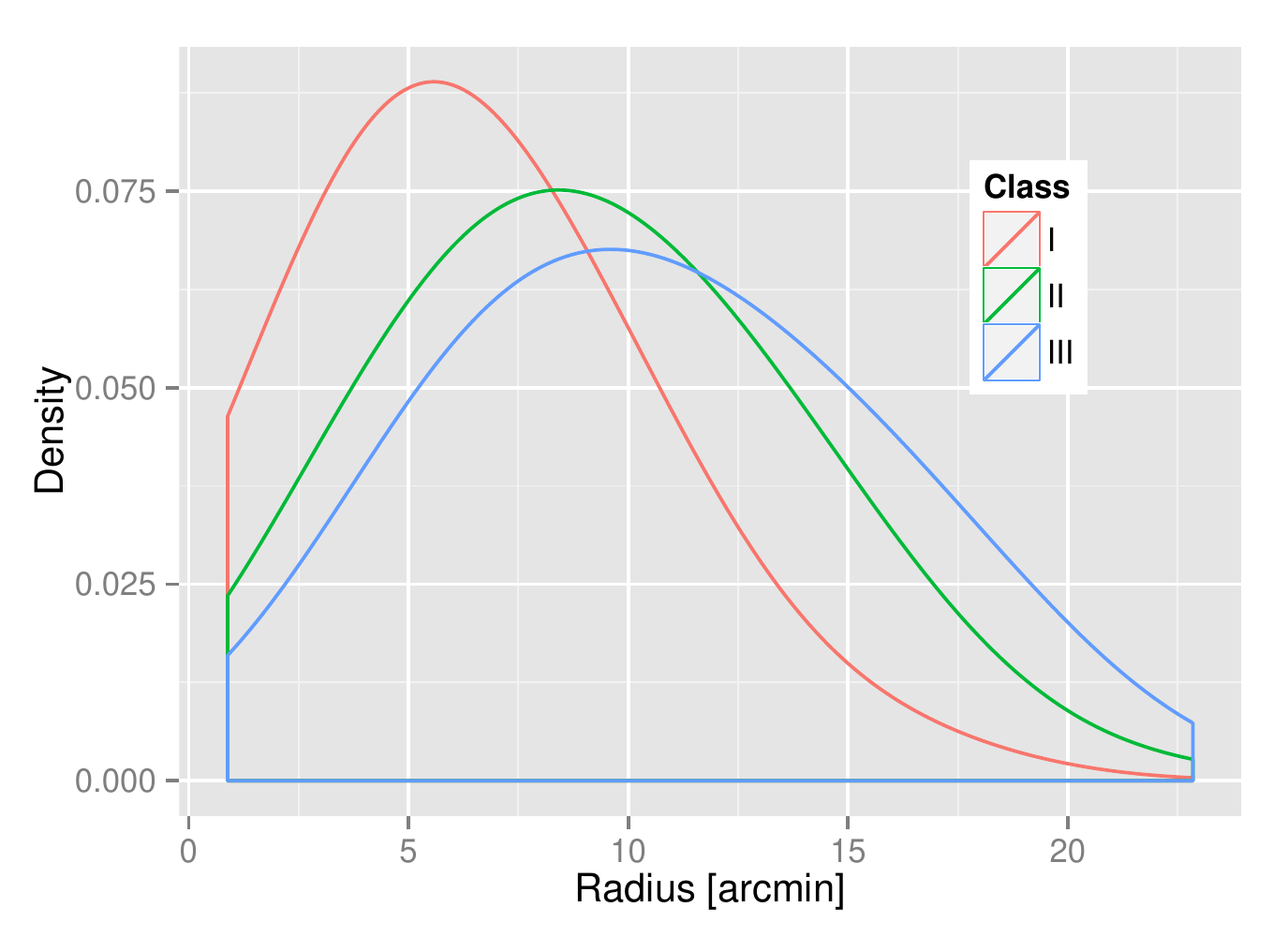}	
\caption{\label{density}Radial density plot. The projected densities of each class
of YSOs are given in number of objects par arcmin$^2$ and are displayed versus the projected distance to the center [of
each class.}
\end{center}
\end{figure}

To analyse how probable it was to find a young stellar object of classes I, II, or III at a given distance from the center of the $\rho$ Ophiuchi cluster F core, we built probability density estimations using kernel density estimators. The  projected radii for the objects classified as members of each SED class were computed based on the angular distance between the positions of each object and the median position of all member objects (Fig.\ref{density}). The median projected radii and the absolute median deviation for each type of object shows that the different classes present different preferential projected radii and also different spread: $R_{ I} = 5.6 \pm 3.1$ \arcmin, $R_{II} = 8.8 \pm 4.5$ \arcmin, $R_{III} = 10.7 \pm 5.3$ \arcmin . These results indicate that the youngest objects are more concentrated in the central parts of the working zone (cores E and F), while more evolved objects are more spread towards the exterior.

\begin{figure}[!htp]    
\begin{center}
\includegraphics[width=0.49\textwidth]{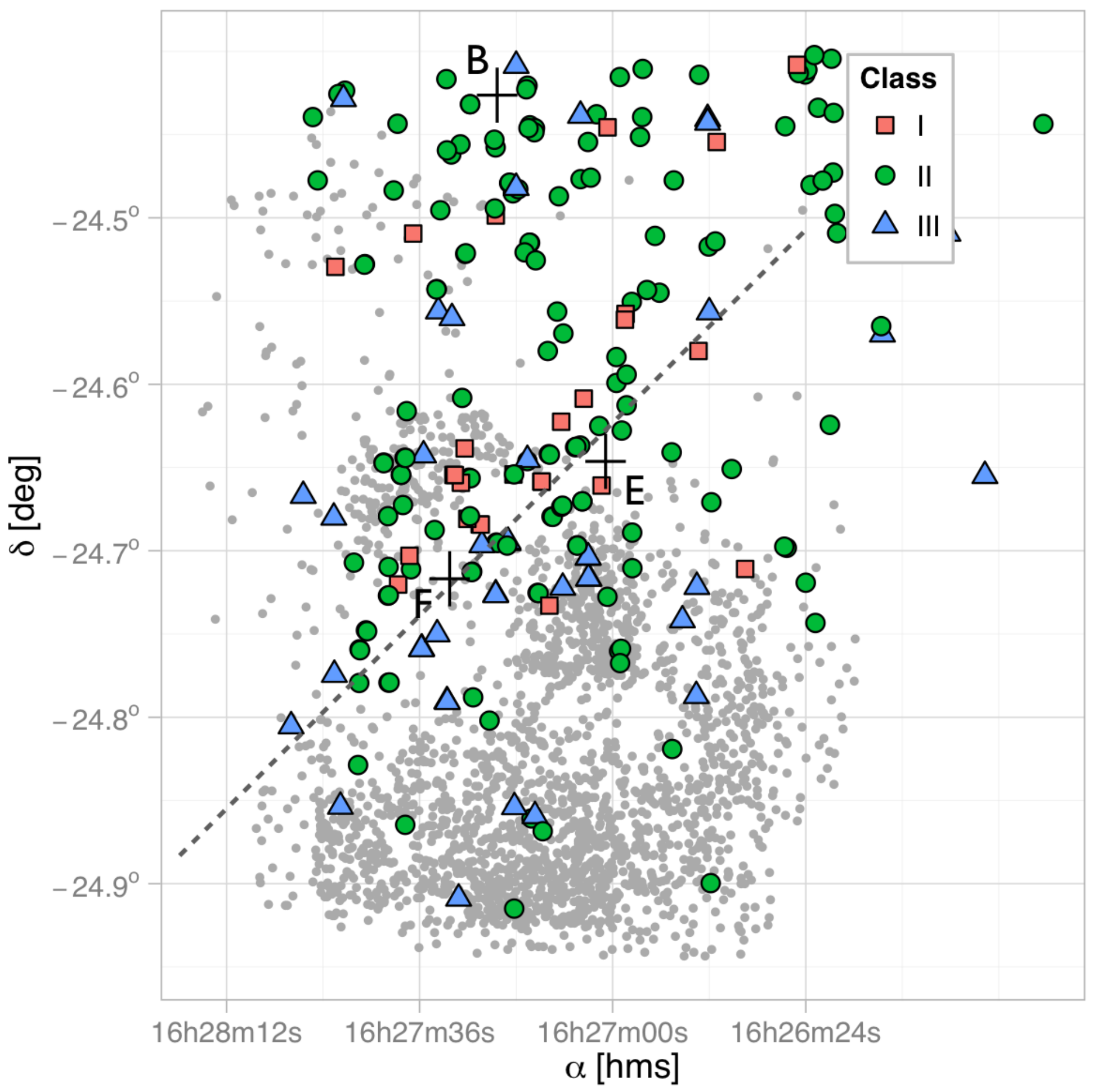}	
\caption{\label{pos} Distribution on the sky of objects from our proper motions catalogue (grey dots) and of the 82 kinematic \textit{members} and \textit{candidates} of the $\rho$ Ophiuchi cluster (coloured symbols) derived in this work. Red squares designate class I, green circles class II and blue triangles class III. The zones with few grey dots correspond to the most obscure regions. Crosses indicate the approximative centres of the E, F, and B cores; the dotted line indicates the line of clumps and YSOs identified by \cite{Motte(1998)}.}
\end{center}
\end{figure}

This is also visible in Fig.\ref{pos}, which displays the distribution on the sky of the objects of our proper motion catalogue and of the kinematic members. We observe that they are mostly located in the region of highest absorption where star-formation is ongoing. We also note that class I objects are globally aligned on a northwest-southeast axis, mostly located in the most embedded region, which was proposed by \cite{Motte(1998)} to correspond to the perpendicular to the direction of shock propagation by the Upper Scorpius OB association onto the core of the $\rho$ Oph cloud complex.   

\section{Nature of kinematic members and candidates}
Of the 82 kinematic members and candidate members in the region we analysed, we proposed a classification for 53 objects. Six objects (2MASS-6X J16271965-2441487, 2MASS J16272439-2441475, 2MASS J16272661-2440451, 2MASS J16273272-2445004, 2MASS J16274161-2446447, and 2MASS J16273894-2440206) have been listed by \cite{Alves(2010)} as potential brown dwarfs (BDs) based on their position in a colour-magnitude diagram among the faintest, red objects. That we can kinematically confirm that they are part of the cluster indicates that they are probably BDs. 

\begin{figure}[!htp]    
\begin{center}
\includegraphics[width=0.49\textwidth]{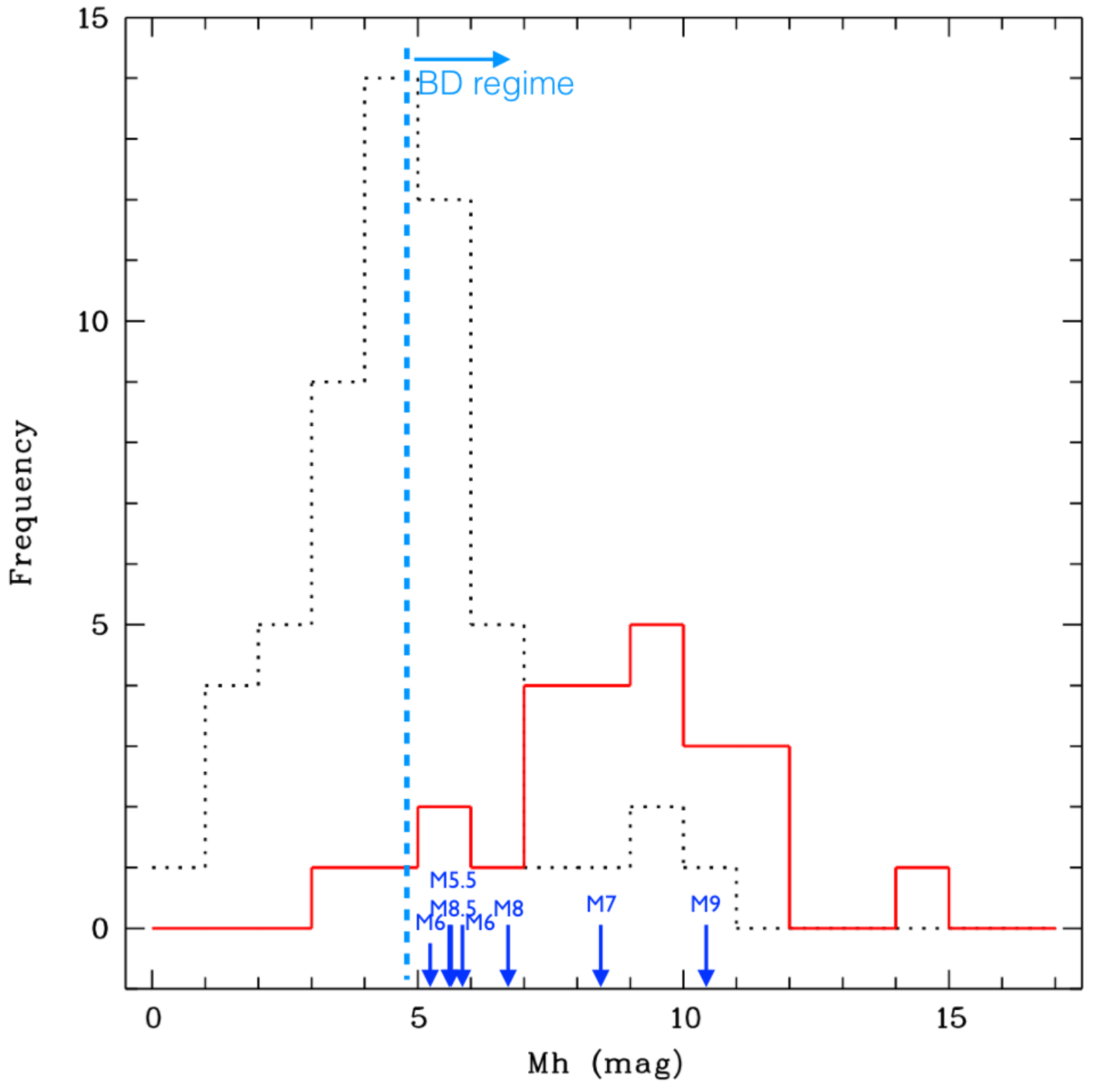}	
\caption{\label{MH} Distribution of absolute H magnitudes of known (black dotted) and new (solid red) members. Spectroscopically confirmed brown dwarfs by \cite{Alves(2012)} are indicated together with their spectral type.}
\end{center}
\end{figure}

Using the typical IR excesses in J, H, and K bands for class II and class III YSOs and the extinction law in near-IR as in \cite{Bontemps(2001)}, we derived the absolute J and H magnitudes for the kinematic members and candidates members\footnote{One new member (WISE J162702.05-243938.6) was detected in only one near-IR band therefore we were unable to derive its absolute magnitudes.}. We present the distribution of these H absolute magnitudes for known members (black dotted histogram) and new members (red) in Fig. \ref{MH} . 

As expected, most of the new members are weak sources with M$_{H}$ ranging from 5 to 12. Since the transition between stars and BDs occurs at M$_{H}$$\sim$5 mag, that is, M$_{J}$$\sim$6 mag for young clusters (e.g., \citealp{Baraffe(2015)}), virtually all the new members fall in the substellar regime. The M$_{H}$ values of the seven spectroscopically confirmed BDs in $\rho$ Ophiuchi are also indicated in Fig \ref{MH} together with their spectral type. Above M$_{H}$ = 5 mag, we find 23 new substellar members, which is 3.3 times more than previously known BDs in the surveyed area.

The weakest four new members (UGCS J162746.78-244059.1, BX162727602-24381862, UGCS J162700.96-244339.5, and 2MASS-6X J16265864-2443281) with M$_{H}$ $\ge$11 mag are potentially very low mass BDs that reach spectral types as cool as L0 ($\sim$2200K, see discussions in \citealp{Tottle(2015)}). These cool young BDs are expected to be very low mass of only about  10 M$_{\rm Jup}$ \citep{Chabrier(2000)}. 

We summarise the evolutionary status of the 82 \textit{members} and \textit{candidates} in Table \ref{class}. Seven members already known as YSOs are identified BD \citep{Alves(2012)} and 6 new YSOs are BD candidates \citep{Alves(2010)}.

\begin{table}[!ht]
\caption{\label{class}Summary of evolutionary status of the 82 kinematic \textit{members} and \textit{candidates}. }
\begin{tabular}{lrr}
\hline
Status		& Members	& Candidates	\\
\hline
56 known YSOs (CDS)  & & \\
Class I  & 4   & 0  \\
Class II &34 & 2 \\
Class III &13 & 3 \\
\hline
26 new YSOs&&\\
Class I  & 2   & 3 \\
Class II &11 & 5\\
Class III &4 & 1\\                        
\hline
  \end{tabular}
\end{table}

\section{Conclusions}
With our astrometric observations we have determined the proper motions of 2213 stellar and sub-stellar objects in the $\rho$~Ophiuchi cluster region. We performed a kinematic membership analysis and derived a list of 82 kinematic \textit{members} and \textit{candidate} members, 26 of them are new members.  We established in a reliable way the mean kinematic properties of the L1688 dark cloud ($\mu_{\alpha}\cos\delta$, $\mu_{\delta}$)=(-8.2, -24.3)$\pm$0.8 mas/yr and confirm that the velocity of this core is very similar to the one of the Upper Scorpius association.

We assigned a SED class to the 53 unclassified kinematic members or candidates using a non-parametric random forests supervised method for classifying objects using any combination of the (J, H, K, w1, w2, w3 and w4) 2MASS and AllWISE colours. Nine objects are defined as class I, 52 as class II, and 21 objects as class III. 

We discovered 23 new BDs candidates as part of the cluster, potentially multiplying the number of known BDs in $\rho$ Ophiuchi cluster by 3.3. A few of them might be extremely low mass BDs in the 10 M$_{\rm Jup}$ regime.

We were able to establish a secondary astrometric reference frame in the NIR in a region where reference stars in the large existing astrometric catalogues can hardly be found. The Gaia catalogue will undoubtedly be a valuable tool for studying associations of young stars, as Hipparcos was in its time. For a clear view of the star-formation processes and to access the youngest clusters (< a few Myr), infrared astrometric catalogues will be the crucial tool, however. The future space infrared astrometric mission Jasmine \citep{Jasmine(2005)} may in this respect be the first to bring insight into embedded star-forming regions. The secondary reference frame established in this work is a first step on this way.

\begin{acknowledgements} 
We are grateful to E. Moraux and C. Alves de Oliveira for providing us with CFHT observational material.

We acknowledge partial financial support from the French organisation COFECUB and the Brazilian organisation FAPESP and CAPES.

This research has made use of the SIMBAD database, operated at CDS, Strasbourg, France.

This publication makes use of data products from the Wide-field Infrared Survey Explorer, which is a joint project of the University of California, Los Angeles, and the Jet Propulsion Laboratory/California Institute of Technology, funded by the National Aeronautics and Space Administration.

This publication makes use of data products from the Two Micron All Sky Survey, which is a joint project of the University of Massachusetts and the Infrared Processing and Analysis Center/California Institute of Technology, funded by the National Aeronautics and Space Administration and the National Science Foundation.
 
\end{acknowledgements}

\bibliographystyle{aa}
\bibliography{references}


\begin{landscape}
\begin{table} 
\caption{\label{elim}  Astrometric parameters of a list of outliers with a high membership probability ($P$>0.9) excluded from members because their proper motion is beyond 2.5$\sigma$ of the mean proper motion of the cluster.}
\begin{tabular}{lrrrrrrrrrrrrl}
\hline
\hline
Object    &  RA  &   Dec  &    $\mu_{\alpha*}$    &    $\mu_{\delta}$   &  2MASS K$s$   &  Prob & \\
               &  [h m s] &[$^{\circ}$ $^\prime$ $^\prime$$^\prime$]  & [$mas/yr$]   & [$mas/yr$]   &[mag]  &   \\
\hline
2MASSJ16265636-2441204 & 16 26 56.35 & -24 41 20.37 &  -0.740 $\pm$ 1.27& -40.640 $\pm$ 1.18 & 12.907 & 1.00 \\
2MASSJ16265904-2435568 & 16 26 59.05 & -24 35 56.92 &   5.400 $\pm$ 1.59& -28.290 $\pm$ 2.19 &  9.993 & 0.99 \\
2MASSJ16271877-2456063 & 16 27 18.77 & -24 56 06.42 &   5.090 $\pm$ 2.81& -23.170 $\pm$ 2.84 & 14.910 & 0.97 \\
2MASSJ16272197-2429397 & 16 27 21.98 & -24 29 39.76 &   6.110 $\pm$ 2.84& -25.770 $\pm$ 2.16 &  0.000 & 0.99 \\
2MASSJ16272463-2429353 & 16 27 24.63 & -24 29 35.39 &   7.190 $\pm$ 2.30& -27.350 $\pm$ 2.67 &  0.000 & 0.99 \\
BX162725643-24372840   & 16 27 25.64 & -24 37 28.41 & -16.750 $\pm$ 0.99& -34.760 $\pm$ 1.12 & 13.996 & 1.00 \\
2MASSJ16272693-2440508 & 16 27 26.93 & -24 40 50.87 & -22.240 $\pm$ 0.67& -22.350 $\pm$ 0.88 &  5.007 & 1.00 \\
UGSC162728.13-243719.5 & 16 27 28.14 & -24 37 19.59 & -17.450 $\pm$ 2.38& -16.610 $\pm$ 2.49 &  0.000 & 0.92 \\
2MASSJ16273213-2429435 & 16 27 32.14 & -24 29 43.59 & -10.360 $\pm$ 1.93& -36.750 $\pm$ 2.60 & 10.740 & 1.00 \\
2MASSJ16273724-2442380 & 16 27 37.24 & -24 42 38.01 &   4.580 $\pm$ 1.27& -21.260 $\pm$ 1.29 &  7.856 & 0.92 \\
2MASSJ16275191-2446296 & 16 27 51.92 & -24 46 29.54 & -13.080 $\pm$ 1.95& -36.840 $\pm$ 2.02 &  9.314 & 1.00 \\
\hline
\hline
\end{tabular}
\tablefoot{$\mu_{\alpha*}$ stands for $\mu_{\alpha}cos(\delta)$. Objects not detected by 2MASS or WISE were searched in the 2MASS 6X  \citep{Cutri(2012)} and the UKIDSS \citep{Lawrence(2007)} catalogues and attributed their identificators. A local identificator (BXhhmmss.sss+dddmmss.ss) was attributed to the faintest source not found in these catalogues. }
\end{table}
\end{landscape} 

\begin{landscape}
\begin{table}[!htp] 
\caption{\label{photo} Extract of Table - Photometry given by 2MASS (J,H,K)  and Allwise (w1,w2,w3,w4) for the kinematic members (as determined in Sect. \ref{membership}) of the $\rho$ Ophiuchi cluster. Values preceded by an asterisk were derived in this work (for J,H,Ks) and come from WISE \citep{Wright(2010)} (for w1,w2, w3, w4).}

\begin{tabular}{lrrrrrrrrrrrrrr}
\hline
Name & J &$\sigma_{J}$ & H &$\sigma_{H}$&K$s$&$\sigma_{Ks}$& w1&$\sigma_{w1}$& w2 &$\sigma_{w2}$ &w3 &$\sigma_{w3}$&w4&$\sigma_{w4}$\\
\hline
2MASS J16264172-2453586 & 14.791 & 0.039 & 14.104 & 0.039 & 13.740 & 0.045 & 13.604 & 0.032 & 13.293 & 0.041 & 11.997 & 0.490 &  7.866 &      \\
2MASS J16264441-2447138 & 11.833 & 0.022 & 10.997 & 0.022 & 10.635 & 0.021 & 10.417 & 0.023 & 10.219 & 0.021 &  9.462 & 0.100 &  7.414 & 0.196\\
2MASS J16264890-2449087 & 16.306 & 0.100 & 14.846 & 0.072 & 14.206 & 0.068 & 13.789 & 0.058 & 13.556 & 0.070 & 10.614 &       &  7.212 & 0.388\\
2MASS-6X J16265634-2442375 & 17.772 & 0.188 & 16.591 & 0.142 & 15.759 & 0.131 &        &       &        &       &        &       &        &      \\
2MASS J16265843-2445318 & 10.365 & 0.030 &  8.635 & 0.090 &  7.549 & 0.046 &  6.991 & 0.120 &  5.984 & 0.096 &  3.557 & 0.059 &  1.217 & 0.069\\
2MASS J16265861-2446029 & 15.353 & 0.047 & 13.966 & 0.035 & 13.314 & 0.027 &        &       &        &       &        &       &        &      \\
UGCS J162700.96-244339.5 &*19.456 & 0.000 &*17.945 & 0.037 &*17.066 & 0.000 &        &       &        &       &        &       &        &      \\
\hline
\hline
\end{tabular}
\tablefoot{Objects not detected by 2MASS or WISE were searched in the 2MASS 6X \citep{Cutri(2012)} and the UKIDSS \citep{Lawrence(2007)} catalogues and attributed their identificators. A local identificator (BXhhmmss.sss+dddmmss.ss) was attributed to the faintest source not found in these catalogues. }
\end{table}
\end{landscape}

\begin{landscape}
\begin{table}[ht] 
\caption{\label{members} Extract of Table - Astrometric parameters of the kinematic members derived in this work with their associated membership probability. Colums. (7) and(8) provide the mean epoch corresponding to the J2000 RA, DEC positions, Col. (10) lists the number of epochs of the observations, Col. (11) their time base and Col. (12) the membership probability.}
\begin{tabular}{lrrrrrrrrrrrrl}
\hline
\hline
Object    &  RA  &   Dec  &   $\sigma_{pos}$   &    $\mu_{\alpha*}$    &    $\mu_{\delta}$   &  EpRA  &  EpDEC  &  2MASS K$s$   &   $N_{Ep}$   &    Dt   &  Prob & \\
                                                 &  [h m s]    &    [$^{\circ}$ $^\prime$ $^\prime$$^\prime$]  &  & [$mas$] & [$mas/yr$]   & [$yr$] &  [$yr$]  & [mag]  &  &  [yr]  &  \\
\hline
2MASS J16264172-2453586 &  16 26 41.7218 & -24 53 58.618 & 5 & -11.8 $\pm$ 0.7 & -21.4 $\pm$ 1.4 & 2007.9 & 2007.8 & 13.740 $\pm$ 0.045 &  6 & 29.6 & 0.99\\
2MASS J16264441-2447138 &  16 26 44.4164 & -24 47 13.835 & 6 &  -4.8 $\pm$ 2.1 & -23.2 $\pm$ 2.2 & 2006.7 & 2006.7 & 10.635 $\pm$ 0.021 &  7 & 29.6 & 1.00\\
2MASS J16264890-2449087 &  16 26 48.9022 & -24 49  8.673 & 4 & -13.4 $\pm$ 1.5 & -23.0 $\pm$ 1.3 & 2007.0 & 2007.0 & 14.206 $\pm$ 0.068 &  5 & 12.1 & 1.00\\
2MASS-6X J16265634-2442375 &  16 26 56.3364 & -24 42 37.616 & 3 & -13.2 $\pm$ 1.3 & -25.1 $\pm$ 1.1 & 2007.2 & 2007.2 & 15.759 $\pm$ 0.131 &  4 &  5.4 & 1.00\\
2MASS J16265843-2445318 &  16 26 58.4414 & -24 45 31.878 & 6 & -11.3 $\pm$ 1.7 & -25.1 $\pm$ 1.9 & 2007.6 & 2007.6 &  7.549 $\pm$ 0.046 &  6 &  9.9 & 1.00\\
2MASS J16265861-2446029 &  16 26 58.6166 & -24 46  2.864 & 2 &  -6.3 $\pm$ 0.9 & -20.0 $\pm$ 0.8 & 2007.4 & 2007.5 & 13.314 $\pm$ 0.027 &  8 & 12.1 & 0.97\\
UGCS J162700.96-244339.5 &  16 27  0.9758 & -24 43 39.530 & 2 & -17.7 $\pm$ 0.8 & -26.5 $\pm$ 0.6 & 2007.6 & 2007.7 &          &  5 &  6.1 & 1.00\\
\hline
\hline
\end{tabular}
\tablefoot{$\mu_{\alpha*}$ stands for $\mu_{\alpha}cos(\delta)$. Note that K$s$ magnitudes without error are just estimative. Objects not detected by 2MASS or WISE were searched in the 2MASS 6X  \citep{Cutri(2012)} and the UKIDSS \citep{Lawrence(2007)} catalogues and attributed their identificators. A local identificator (BXhhmmss.sss+dddmmss.ss) was attributed to the faintest source not found in these catalogues. }
\end{table}
\end{landscape} 

\begin{landscape}
\begin{table}[ht] 
\caption{\label{classification} Extract of Table - Evolutionary status of the kinematic members. Columns (1) and (2) provide identifications of the objects. Column (3) (status) lists the young stellar object status as given by CDS-Simbad and Column (4) (Class) the SED class of the object (classes preceded by an asterisk were assigned in this work). The corresponding references are given in Column (5) .  In Column (6) we list the spectral type found in the litterature and in Col. (7) the references. }
\begin{tabular}{lllrrrrrrr}
\hline
\hline
Object                                      & Other Id.             &  Position                            &  Status& Class & ref. & SpT & ref.           \\
\hline
2MASS J16264172-2453586  &                             &  16 26 41.721 -24 53 58.61   & 	& *II  &    &	    &	 \\
2MASS J16264441-2447138  &SSTc2d J162644.4-244714      &  16 26 44.416 -24 47 13.83   & TT*	& *III &    & M4.5  &  2 \\
2MASS J16264890-2449087  &                             &  16 26 48.902 -24 49 08.67   & 	& *II  &    &	    &	 \\
2MASS-6X J16265634-2442375  &                             &  16 26 56.336 -24 42 37.61   & 	& *II  &    &	    &	 \\
2MASS J16265843-2445318  &YLW 3A                       &  16 26 58.441 -24 45 31.87   & TT*	& *II  &    & M0-K8 & 10 \\
2MASS J16265861-2446029  &                             &  16 26 58.616 -24 46 02.86   & 	& *II  &    &	    &	 \\
UGCS J162700.96-244339.5  &                             &  16 27 00.975 -24 43 39.53   & 	& *II  &    &	    &	 \\
\hline
\hline
\end{tabular}
\tablefoot{
Objects not detected by 2MASS or WISE were searched in the 2MASS 6X \citep{Cutri(2012)} and the UKIDSS \citep{Lawrence(2007)} catalogues and attributed their identificators. A local identificator (BXhhmmss.sss+dddmmss.ss) was attributed to the faintest source not found in these catalogues. a: this object is classified as class II by \cite{Bontemps(2001)} but was classified as class III by our algorithm. b: this object is classified as class I by \cite{Bontemps(2001)} but was classified as class II by our algorithm.}
\tablebib{
(1)~\cite{Muzic(2012)};
(2)~\cite{Wilking(2008)}; 	     
(3)~\cite{Bontemps(2001)};
(4)~\cite{Cieza(2007)};
(5)~\cite{Pillitteri(2010)};
(6)~\cite{Wilking(2005)};
(7)~\cite{Ozawa(2005)};
(8)~\cite{Erickson(2011)};
(9)~\cite{Alves(2008)};
(10)~\cite{Martin(1998)};
(11)~\cite{Luhman(1999)};
(12)~\cite{Simon(1995)};
(13)~\cite{Gutermuth(2009)};
(14)~\cite{Parks(2014)};
(15)~\cite{Natta(2002)};
(16)~\cite{Andrews(2007)};
(17)~\cite{Furlan(2009)};
(18)~\cite{Natta(2006)};
(19)~\cite{Alves(2012)};
(20)~\cite{Dodson(2011)};. 
(21)~\cite{Rodriguez(2013)};	
(22)~\cite{Gutermuth(2009)};	

}\end{table}
\end{landscape}

\end{document}